\documentclass{llncs}
\usepackage{graphicx}
\usepackage{floatflt}
\usepackage{wrapfig}
\usepackage{makeidx}  
\usepackage{float}
\restylefloat{table}
\usepackage{booktabs}
\usepackage{multirow}
\usepackage{algorithm}
\usepackage[noend]{algpseudocode}
\algnewcommand{\IIf}[1]{\State\algorithmicif\ #1\ \algorithmicthen}
\algnewcommand{\EndIIf}{\unskip\ }
\makeatletter

\algrenewcommand\ALG@beginalgorithmic{\small}
\makeatother
\algrenewcommand\alglinenumber[1]{\scriptsize #1:}

\makeatletter
\let\OldStatex\Statex
\renewcommand{\Statex}[1][3]{%
  \setlength\@tempdima{\algorithmicindent}%
  \OldStatex\hskip\dimexpr#1\@tempdima\relax}
\algnewcommand{\LineComment}[1]{\OldStatex \hskip\ALG@thistlm \(\triangleright\) #1}

\algrenewcommand\algorithmicindent{1.0em}%
\newcommand{\getmin}{{\sc GetMinCut}\xspace}
\newcommand{\getsucc}{{\sc GetSuccessor}\xspace}
\newcommand{\getsuccopt}{{\sc GetSuccessorOptimized}\xspace}

\usepackage{hyperref}
\usepackage{tikz}
\usepackage{xspace}
\usetikzlibrary{calc}
\usetikzlibrary{calc,trees,positioning,arrows,chains,shapes.geometric,fit,
    shapes,shadows,matrix}
\usetikzlibrary{shapes.arrows}
\usetikzlibrary{shapes.geometric,positioning}

\usepackage{amssymb}
\usepackage{amsmath}
\usepackage{epstopdf}
\usepackage{subfig}
\usepackage{caption}
\newcommand{\cuts}[1]{\mathcal{C}(#1)}
\newcommand{\bigo}[1]{\mathcal{O}(#1)}
\newcommand{\nuni}{$n_u$\xspace}

\newcommand{\Ra}{\Rightarrow}

\newcommand{\ra}{\rightarrow}

\newcommand{\figref}[1]{Fig.~\ref{#1}}
\newcommand{\algoref}[1]{Algorithm~\ref{#1}}

\graphicspath{{figures/}}

\newtheorem{t1}{Theorem}
\newtheorem{l1}{Lemma}

\newtheorem{d1}{Definition}

\newcommand{\C}[2]{{#1} \choose {#2}}

\newcommand{\remove}[1]{}

\begin{document}






%

\title{Space Efficient Breadth-First and Level Traversals of Consistent Global States of Parallel Programs}
\author{Himanshu Chauhan \and Vijay K. Garg}
\institute{University of Texas at Austin\\
\email{himanshu@utexas.edu ~~~~ garg@ece.utexas.edu} }

\date{}

\maketitle
\begin{abstract}
Enumerating {\em consistent} global states of a  
 computation is a fundamental problem in parallel
computing with applications to debugging, testing and runtime verification of parallel programs.
Breadth-first search (BFS) enumeration is especially useful for these applications as it finds an erroneous
consistent global state with the least number of events possible. The total 
number of executed events in a global state is called its {\em rank}. 
BFS also allows enumeration of all 
 global states of a given rank or within a range of ranks.
 \remove{
 were the first to give an algorithm for BFS enumeration of consistent global states.
Their algorithm stores all global states of  rank $r-1$ 
to enumerate global states of rank $r$. 
In the worst case, this storage can be  
 exponential in the number of processes.   
 We give the first algorithm that enumerates consistent global states 
 of a given rank without enumerating states of lower rank. This 
 enables BFS based enumeration of  
consistent global states with polynomial space 
in the number of processes.} If a computation on $n$ processes has 
$m$ events per process on average, then the traditional BFS (Cooper-Marzullo 
and its variants) requires
$\bigo{\frac{m^{n-1}}{n}}$ space in the worst case, whereas our algorithm performs the BFS 
requires  $\bigo{m^2n^2}$ space. Thus, we reduce the 
space complexity for BFS enumeration of consistent global states  exponentially.  
and give the first polynomial space algorithm for this task.   
In our experimental evaluation of seven benchmarks,   
 traditional BFS fails in many cases by exhausting the 2 GB heap space allowed to the JVM. 
In contrast, our implementation uses less than 60 MB memory 
and is also faster in many cases. 
\end{abstract}
%
%
%
%

%
%


\section{Introduction} 
\label{sec:intro}

Parallel programs are not only difficult to 
design and implement, but once implemented are also difficult
to debug and verify. 
The technique of predicate detection \cite{cite:GW94,CoopMarz:ConsDetGP} is helpful 
in verification of these implementations as it allows inference  
based analysis to check many possible system states based on 
one execution trace. The technique involves execution of the program, 
and modeling of its trace as a partial order. Then all possible 
states of the model that are consistent with the partial order are 
visited and evaluated for violation of any constraints/invariants. 
A large body of work uses this approach to verify distributed applications, 
as well as 
to detect  data-races and 
 other concurrency related bugs in shared memory 
 parallel programs \cite{cite:jpredictor,cite:fasttrack,cite:pecan,cite:atomicityviolation}.  
 Finding consistent global states of an execution also has critical applications in
snapshotting of modern  
distributed file systems \cite{AlagappanEtAl16-OSDI,theo}.

 A fundamental requirement for this approach is the traversal of all 
 possible consistent global states, or {\em consistent cuts}, of a parallel
 execution.  
Let us call the execution of a parallel program a {\em computation}.
\remove{ 
that comprises of a set of events or operations $E$, 
such that each event is executed by  one of the   
  $n$ parallel processes, $P_1$ to $P_n$, involved in the execution. } 
The set of all consistent cuts of a computation can be represented 
as a directed acyclic graph in which each vertex  
represents a consistent cut, and the edges mark the transition from 
one global state to another by executing one
 operation. 
 Moreover, this graph has a special structure: it is a distributive 
lattice \cite{Mat:1989:WDAG}.
\remove{
any two vertices (consistent cuts) in the graph have a unique 
 union and a unique intersection; and 
 there is only one 
source vertex, and only one sink vertex. }
Multiple algorithms have been proposed to traverse 
the lattice of consistent cuts of a parallel execution.     
Cooper and Marzullo's algorithm\cite{CoopMarz:ConsDetGP}
starts from the source --- a consistent cut in which no operation 
has been executed by any process --- and performs a breadth-first-search (BFS)
visiting the lattice level by level.  
Alagar and Venkatesan's algorithm\cite{AlaVen:1993:WAD}
performs a depth-first-search (DFS) traversal of the lattice, and
Ganter's algorithm \cite{cite:Ganter10} enumerates global states in lexical order.
\remove{ and
requires $\bigo(nM)$ time and $\bigo(nE)$ space, where 
$M$ is the number of
consistent cuts. 
The main disadvantage of their
algorithm is that it requires recursive calls of depth
$\bigo(E)$ with each call requiring $\bigo(n)$ space resulting
in $\bigo(nE)$ space requirements besides storing the computation
itself. 
Ganter \cite{cite:Ganter10} presented an algorithm, which enumerates global states in lexical order, and Garg \cite{cite:enumeratingglobal} gave an implementation using vector clocks \cite{cite:vectorclock1,cite:vectorclock2}.
The lexical algorithm requires $\bigo(n^2)$ time, but the algorithm requires no additional space besides the input, i.e., the computation.
Chang and Garg \cite{ChangGarg15} reduced
the time complexity from $\bigo(n^2)$ to $\bigo(n \cdotp \Delta(P))$
where $\Delta(P)$ is the maximal in-degree of any event. }

The BFS traversal of the lattice is particularly useful in solving 
two key problems. First, 
suppose a programmer is debugging a parallel program to find a 
concurrency related bug. The global state in which this bug 
occurs
is a counter-example to the programmer's understanding of a correct execution, 
and we want to halt the execution of the program on reaching the first 
state where the bug occurs. 
Naturally, finding a small counter example is quite useful in such cases. 
The second problem 
 is to check all consistent cuts of given rank(s). 
 For example, a programmer may observe that her program crashes only after $k$ 
 events have been executed, or while debugging an implementation 
of Paxos \cite{lamport2001paxos} algorithm, she might only be 
interested in analyzing the system when all processes have sent their {\em promises} 
to the leader.     
Among the existing traversal algorithms,  
 the BFS algorithm provides a straightforward solution to these two problems.
It is guaranteed to traverse the lattice of consistent cuts 
in a level by level manner where each level corresponds to 
the total number of events executed in the computation.       
This traversal, however, requires space
proportional to the size of the biggest level of
the lattice which, in general, is {\em exponential} in the size
of the computation.

In this paper, we present a new algorithm
to perform BFS traversal of the lattice
in space that is polynomial in the size of the computation. In short, the contribution 
of this paper are: 
\begin{itemize}\itemsep0em
   \item For a computation on $n$ processes such that each process has $m$ 
   events on average, our algorithm requires $\bigo{m^2n^2}$ 
   space in the worst case, 
   whereas the traditional BFS algorithm requires $\bigo{\frac{m^{n-1}}{n}}$ space  (exponential in 
   $n$). 
   \item Our evaluation on 
seven benchmark computations shows the traditional BFS runs out of the maximum allowed 2 GB memory for three of them, whereas our implementation can traverse the lattices by using less than 60 MB memory for each 
benchmark.
\end{itemize}
 \remove{
Consider the execution  on two processes $P_1$ and $P_2$ shown in 
\figref{fig:vecclocks}. Each process executes three events, and 
event $f$ on $P_2$ is causally dependent on event $b$ on $P_1$. This 
execution has fourteen possible consistent global states in all.  
}

The exponential reduction in space is sometimes at the cost of a loss in time required to perform the BFS traversal. Our analysis in experimental results

\section{Background}
\label{sec:model}

We model a computation $P = (E,\ra)$ on 
$n$ processes $\{P_{1}, P_{2}, 
\ldots, P_{n} \}$ as a partial order
 on the set of events, $E$. The events are ordered by Lamport's \emph{happened-before} ($\ra$) relation 
\cite{Lam:1978:CACM}.
This partially ordered set (poset) of events is partitioned into 
chains:
\begin{d1}[Chain Partition]
A chain partition of a poset places every element of the 
poset on a chain that is totally ordered. 
Formally, if $\alpha$
 is a chain partition of poset $P = (E,\ra)$ then $\alpha$ maps every event 
to a natural number such that 
\[ \forall x,y \in E: \alpha(x) = \alpha(y) \Ra (x \ra y) \vee (y \ra x). \]
\end{d1}

Generally, a computation on $n$ processes is partitioned into $n$ chains 
such that the events 
executed by process $P_i$ ($1 \leq i \leq n$) are placed on 
$i^{th}$ chain. 

Mattern \cite{Mat:1989:WDAG} and Fidge \cite{Fid:1988:ACSC} proposed {\em vector clocks}, an approach for time-stamping events in a computation such that 
the happened-before relation can be tracked. For a program 
on $n$ processes, each event's vector clock is a $n$-length vector of 
integers. Note that vector clocks are dependent on chain 
partition of the poset that models the computation. For an event $e$, we denote $e.V$ as its vector clock. 

Throughout this paper, we use the following 
representation for interpreting chain partitions and vector clocks:  
if there are $n$ chains in the chain partition of the computation, 
then the lowest chain (process) is always numbered $1$, and the highest chain being numbered $n$. 
 A vector clock on $n$ chains is represented as a $n$-length vector: 
$[c_n,c_{n-1},...,c_i,...,c_2,c_1]$ such that $c_i$ denotes 
the number of events executed on process $P_i$.   

Hence, if event $e$ was executed on process $P_i$, then $e.V[i]$ is $e$'s index (starting from $1$) on $P_i$. Also, for any event $f$ in
the computation: $
e \ra f \Leftrightarrow \forall j: e.V[j] \leq f.V[j] \land \exists k: e.V[k] < f.V[k]$.
A pair of events, $e$ and $f$, is concurrent iff $e \not \ra 
f \wedge f \not \ra e$.
We denote this relation by $e || f$. 
\remove{If $f$ is executed by $P_j$, 
and $e || f$, then we must have that  $e.V[i] > f.V[i] \land e.V[j] < f.V[j]$. }
Fig.~\ref{fig:vecclocks} shows a 
sample computation with six events and their corresponding 
vector clocks.  Event $b$ is the second event on process $P_1$, 
and its vector clock is $[0,2]$. 
Event $g$ is the third event on $P_2$, but it is preceded by $f$, which 
in turn is causally dependent on $b$ on $P_1$, and thus 
the vector clock of $g$ is $[3,2]$. 
\tikzstyle{place}=[circle,draw=black,fill=black,thick,inner sep=0pt,minimum size=2mm]
\tikzstyle{nsplace}=[circle,draw=black,thick,inner sep=0pt,minimum size=2mm]
\tikzstyle{place1}=[circle,draw=black,thick,inner sep=0pt,minimum size=3mm]
\tikzstyle{inner}=[circle,draw=black,fill=black,thick,inner sep=0pt,minimum size=1.25mm]
\tikzstyle{satlatnode}=[ellipse,draw=black,fill=black!20,thick,minimum size=5mm]
\tikzstyle{nsatlatnode}=[ellipse,style=dashed,thick]
\tikzstyle{satblock} = [rectangle, draw=gray, thin, fill=black!20,
text width=2.5em, text centered, rounded corners, minimum height=2em]
\tikzstyle{nsatblock} = [text width=2em, text centered, minimum height=1em]
\tikzstyle{jbblock} = [rectangle, draw=black, thick, fill=black!20,
text width=2.5em, text centered, minimum height=2em]
\vspace{-2ex}
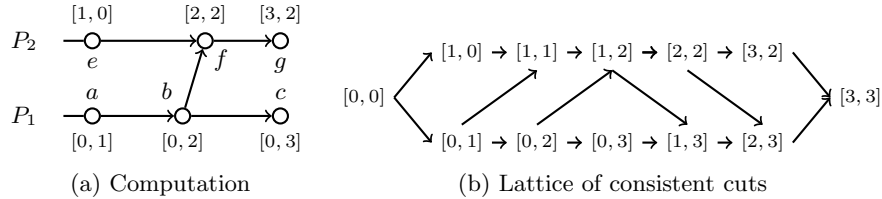
\begin{figure}[htb]
\centering
\subfloat[Computation]{ 
\begin{tikzpicture}
\node at ( 0,1) [nsplace] [label={below:$e$},label={above:{\scriptsize $[1,0]$}}] (a) {};
\node at ( 1.5,1) [nsplace] [label={[xshift=6pt, yshift=-18pt]$f$},label={above:{\scriptsize $[2,2]$}}] (b) {};
\node at ( 2.5,1) [nsplace] [label={below:$g$},label={above:{\scriptsize $[3,2]$}}] (c) {};
\node at ( 0,0) [nsplace] [label={above:$a$},label={below:{\scriptsize $[0,1]$}}] (e) {};
\node at ( 1.2,0) [nsplace] [label={[xshift=-6pt, yshift=0pt]$b$},label={below:{\scriptsize $[0,2]$}}] (f) {};
\node at ( 2.5,0) [nsplace] [label={above:$c$},label={below:{\scriptsize $[0,3]$}}] (g) {};
%
%
\draw [thick,->] (a.east) -- (b.west);
\draw [thick,->] (b.east) -- (c.west);
\draw [thick,->] (e.east) -- (f.west);
\draw [thick,->] (f.east) -- (g.west);
\node at (-0.5, 1) [label=left:$P_{2}$] (bot1) {};
\node at (-0.5, 0) [label=left:$P_{1}$] (bot2) {};

\draw [thick] (a.west) -- (bot1.east);
\draw [thick] (e.west) -- (bot2.east);
\draw [thick,->] (f) -- (b);
\end{tikzpicture}
\label{fig:vecclocks}
}
\subfloat[Lattice of consistent cuts]{ 
\begin{tikzpicture} 
\node (zero) at (-.3,0) [] {\scriptsize $[0,0]$} ;
\node (a) at (1,0.6) [] {\scriptsize $[1,0]$};
\node (b) at (2,0.6) [] {\scriptsize $[1,1]$};
\node (c) at (3,0.6) [] {\scriptsize $[1,2]$};
\node (d) at (4,0.6) [] {\scriptsize $[2,2]$};
\node (e) at (5,0.6) [] {\scriptsize $[3,2]$};

\node (f) at (1,-.6) [] {\scriptsize $[0,1]$};
\node (g) at (2,-.6) [] {\scriptsize $[0,2]$};
\node (h) at (3,-.6) [] {\scriptsize $[0,3]$};
\node (i) at (4,-.6) [] {\scriptsize $[1,3]$};
\node (j) at (5,-.6) [] {\scriptsize $[2,3]$};

\node (end) at (6.3,0) [] {\scriptsize $[3,3]$};

\draw [thick,->] (zero.east) -- (a.west);
\draw [thick,->] (zero.east) -- (f.west);
\draw [thick,->] (a.east) -- (b.west);
\draw [thick,->] (b.east) -- (c.west);
\draw [thick,->] (c.east) -- (d.west);

\draw [thick,->] (f.north) -- (b.south);
\draw [thick,->] (g.north) -- (c.south);
\draw [thick,->] (c.south) -- (i.north);
\draw [thick,->] (d.south) -- (j.north);
\draw [thick,->] (c.east) -- (d.west);
\draw [thick,->] (d.east) -- (e.west);
\draw [thick,->] (f.east) -- (g.west);
\draw [thick,->] (g.east) -- (h.west);
\draw [thick,->] (h.east) -- (i.west);
\draw [thick,->] (i.east) -- (j.west);
\draw [thick,->] (e.east) -- (end.west);
\draw [thick,->] (j.east) -- (end.west);
\end{tikzpicture}
\label{fig:lattice}
}
\caption{A computation with vector clocks of events, and its consistent cuts}
\end{figure}

\begin{d1}[Consistent Cut] 
Given a computation $(E, \ra)$, a subset of events $C \subseteq E$ forms a
\emph{consistent cut} if $C$ contains an event $e$ only if it contains all events that happened-before $e$. Formally,
$
(e \in C) \land (f \ra e) \implies (f \in C)
$.
\end{d1}

A consistent cut captures the notion of a possible global state of the system at some point during
its execution \cite{ChaLam:1985:TrCS}. \remove{We use the term {\em cut} for a possible 
global state of the computation which may or may not be consistent. When the 
global state is consistent, we use the term consistent cut.  }
\remove{
The concept of a consistent cut (or, a consistent global state) is identical to that of a down-set (or 
order-ideal) used in lattice theory \cite{DavPri:1990:CUP}.}
Consider the computation shown in Fig~\ref{fig:vecclocks}. The subset of 
events $\{a, b, e\}$ forms a consistent cut, whereas the subset $\{a, 
e, f\}$ does not; 
because $b \ra f$ ($b$ happened-before $f$) but $b$ is
not included in the subset.\\

\noindent {\bf Vector Clock Notation of Cuts}: 
So far we have described how vector clocks can be used to time-stamp 
 events in the computation. We also use them to   
 represent cuts of the computation. If the computation is 
partitioned into $n$ chains, then for any cut $G$, its vector clock 
 is a $n$-length vector such that $G[i]$ denotes the number 
of events from $P_i$ included in $G$. Note that in our vector 
clock representation the events from $P_i$ are at the $i^{th}$ 
index from the right.

For example, consider the state of the computation in \figref{fig:vecclocks} 
when $P_1$ has executed events $a$ and $b$, and $P_2$ has only executed 
event $e$. The consistent cut for this state, $\{a,b,e\}$, is represented 
by $[1,2]$. Note that  cut $[2,1]$ is not consistent, as it indicates 
execution of $f$ on $P_2$ without $b$ being executed on $P_1$.  
\remove{
\begin{t1}
\cite{DavPri:1990:CUP,Mat:1989:WDAG}
Let $\cuts{E}$ denote the set of all consistent cuts of a computation $(E, \ra)$.
\remove{The set of consistent cuts of any computation $(E, \ra)$}
$\cuts{E}$ forms a finite distributive lattice under the relation $\subseteq$.
\end{t1} 
}
The computation  in Fig.~\ref{fig:vecclocks} has twelve 
consistent cuts; and the lattice of these consistent cuts (in their vector clock 
representation) is shown in Fig.~\ref{fig:lattice}.  

\noindent {\bf Rank of a Cut}: 
Given a cut $G$, we define $rank(G) = \sum G[i]$. The rank of a cut
corresponds to the total number of events, across all processes, that have 
been executed to reach the cut. 

In Fig.~\ref{fig:lattice}, there is one source cut ($[0,0]$) with rank 0, then there are two cuts each
of ranks 1 to 5, and finally there is one cut ($[3,3]$) has rank 6.



\remove{A global 
predicate is \emph{local} if it depends only on variables of a single 
process. 
If a predicate $B$ evaluates to true for a consistent cut $C$, we 
say that ``$C$ satisfies $B$''.\\}

\subsection{Breadth-First Traversal of Lattice of 
Consistent Cuts}

Consider a parallel computation $P = (E,\ra)$. 
The lattice of consistent cuts, $\cuts{E}$, of 
$P$ is a
DAG whose vertices
are the consistent 
cuts of $(E,\ra)$, and there is a directed edge 
from vertex $u$ to vertex $v$ if state represented 
by $v$ can be reached by executing one event 
on $u$; hence we also have 
 $rank(v) = rank(u) + 1$. The source of $\cuts{E}$
is the empty set: a consistent cut in which no events 
have been executed on any process.  
The sink of this DAG is $E$: the consistent cut 
in which all the events of the computation have been 
executed. 
\remove{
Cooper and Marzullo's \cite{CoopMarz:ConsDetGP} 
  algorithm performs breadth-first 
traversal of $\cuts{E}$. Even though they 
focussed on distributed systems, their algorithm 
has been subsequently adopted for verification 
of shared-memory parallel programs too \cite{jpredictor, fasttrack}.  }
Breadth-first search (BFS) of this lattice starts from 
the source vertex and visits all the cuts 
of rank $1$; it then visits all the cuts of rank $2$ 
and continues in this manner till reaching the 
last consistent cut of rank $|E|$. 
For example, in \figref{fig:lattice} the BFS algorithm will traverse cuts in the following order:
 $[0,0], [0,1], [1,0], [0,2], [1,1], [0,3], [1,2], [1,3], [2,2], [2,3], [3,2], [3,3]$. 

The standard BFS on a graph
needs to store the vertices at distance 
$d$ from the source to be able to visit the vertices at
distance $d+1$ (from the source). Hence, in 
performing a BFS 
on $\cuts{E}$ we are required to 
store 
the cuts of rank $r$ in order to visit the 
cuts of rank $r+1$.  Observe that in a parallel 
computation there may 
be exponentially many cuts of rank $r$.\remove{ For example, consider a parallel 
computation on $n$ processes in which no processes 
communicate with each other, and each process 
executes $m$ events individually. For this computation, 
the number of cuts of rank $r \leq m$ is $\C{r.n}{r}$.} Thus, traversing 
the lattice $\cuts{E}$ requires space 
which  is
exponential in the size of input. The optimized vector clock 
based 
 BFS traversal  
takes $\bigo{n^2}$ time per cut \cite{cite:enumeratingglobal}, where 
$n$ is the number of processes in the computation.

\subsection{Related Work}
\label{sec:related}

Cooper and Marzullo \cite{CoopMarz:ConsDetGP} gave the first algorithm for global states enumeration which 
is based on breadth first search (BFS).
Let  $i(P)$ denote the total number 
of consistent cuts of a poset $P$. Cooper-Marzullo 
algorithm requires $\bigo{n^2 \cdot i(P)}$ time, and exponential space in the 
size of the input computation. The exponential space requirement is due to the standard 
BFS approach in which consistent cuts of rank $r$ must be stored to 
traverse the cuts of rank $r+1$. 

There is also a body of work
on enumeration of consistent cuts in order different than BFS.
Alagar and Venkatesan \cite{cite:alagar01} presented a depth first algorithm 
using the notion of global interval which reduces the space complexity to $\bigo{|E|}$. Steiner \cite{cite:steiner86} gave an algorithm that uses $\bigo{|E|\cdot i(P)}$ time, and Squire \cite{cite:Squire95} further improved the computation time to $\bigo{log|E| \cdot i(P)}$.
Pruesse and Ruskey \cite{cite:pruesse93} gave the first algorithm that generates global states in a combinatorial Gray code manner. The algorithm uses $\bigo{|E| \cdot i(P)}$ time and can be reduced to $\bigo{\Delta(P) \cdot i(P)}$ time, where $\Delta(P)$ is the in-degree of an event; however, the space grows exponentially in $|E|$.
Later, Jegou et al. \cite{cite:jegou95} and Habib et al. \cite{cite:habib01} improved the space complexity to $\bigo{n\cdot|E|}$.

Ganter \cite{cite:Ganter10} presented an algorithm, which uses the notion of lexical order, and Garg \cite{cite:enumeratingglobal} gave the implementation using vector clocks. The lexical algorithm requires $\bigo{n^2 \cdot i(P)}$ time but the algorithm itself is {\em stateless} and hence requires no additional space besides the poset. Paramount \cite{chang2015parallel} 
gave a parallel algorithm to traverse this 
lattice in lexical order, and QuickLex \cite{quicklex} 
provides an improved implementation for lexical 
traversal that  
takes $\bigo{n \cdot \Delta(P) \cdot i(P)}$ time, 
and  $\bigo{n^2}$ space overall.  


\section{Uniflow Chain Partition}
\begin{wrapfigure}{R}{0.5\textwidth} 
\begin{minipage}[tbh]{0.5\textwidth}
\vspace{-0.7in}
\tikzstyle{place}=[circle,draw=black,fill=black,thick,inner sep=0pt,minimum size=2mm]
\tikzstyle{nsplace}=[circle,draw=black,thick,inner sep=0pt,minimum size=2mm]
\tikzstyle{place1}=[circle,draw=black,thick,inner sep=0pt,minimum size=3mm]
\tikzstyle{inner}=[circle,draw=black,fill=black,thick,inner sep=0pt,minimum size=1.25mm]
\tikzstyle{satlatnode}=[ellipse,draw=black,fill=black!20,thick,minimum size=5mm]
\tikzstyle{nsatlatnode}=[ellipse,style=dashed,thick]
\tikzstyle{satblock} = [rectangle, draw=gray, thin, fill=black!20,
text width=2.5em, text centered, rounded corners, minimum height=2em]
\tikzstyle{nsatblock} = [rectangle, draw=gray, thin,
text width=2.5em, text centered, rounded corners, minimum height=2em]
\tikzstyle{jbblock} = [rectangle, draw=black, thick, fill=black!20,
text width=2.5em, text centered, minimum height=2em]
\begin{figure}[H]
\centering
\begin{tikzpicture}[scale=0.8, every node/.style={transform shape}]
\node at ( 0,0) [nsplace] [fill=blue!40] (a) {};
\node at ( 1.2,0) [nsplace] [fill=blue!40] (b) {};
\node at ( 2,0) [nsplace] [fill=blue!40] (c) {};
\node at ( 0,-1) [nsplace] [fill=orange!60] (e) {};
\node at ( 1,-1) [nsplace] [fill=orange!60] (f) {};
\node at ( 2,-1) [nsplace] [fill=orange!60] (g) {};
\draw [thick,->] (a.east) -- (b.west);
\draw [thick,->] (b.east) -- (c.west);
\draw [thick,->] (e.east) -- (f.west);
\draw [thick,->] (f.east) -- (g.west);
\draw [thick,->] (f) -- (b);
\node at (-0.2, 0) [label=left:$P_{2}$] (bot1) {};
\node at (-0.2, -1) [label=left:$P_{1}$] (bot2) {};
\node at (1, -1.6) [] (lab) {\normalsize (a)};
\node at ( 3.4,0.6) [nsplace] [fill=blue!40] (a2) {};
\node at ( 4.4,0.6) [nsplace] [fill=blue!40] (b2) {};
\node at ( 5.4,0.6) [nsplace] [fill=blue!40] (c2) {};
\node at ( 3.4,-.2) [nsplace] [fill=orange!60] (e2) {};
\node at ( 4.4,-.2) [nsplace] [fill=orange!60] (f2) {};
\node at ( 5.4,-.2) [nsplace] [fill=orange!60] (g2) {};
\draw [thick,->] (a2.east) -- (b2.west);
\draw [thick,->] (b2.east) -- (c2.west);
\draw [thick,->] (e2.east) -- (f2.west);
\draw [thick,->] (f2.east) -- (g2.west);
\node at (3.3, 0.6) [label=left:$P_{3}$] (bot1) {};
\node at (3.3, -.2) [label=left:$P_{2}$] (bot2) {};
\node at (3.3, -1) [label=left:$P_{1}$] (bot21) {};
\node at (3.4,-1) [nsplace] [fill=gray!60] (a21) {};
\node at (4.4,-1) [nsplace] [fill=gray!60] (b21) {};
\node at (5.4,-1) [nsplace] [fill=gray!60] (c21) {};
\draw [thick,->] (a21.east) -- (b21.west);
\draw [thick,->] (b21.east) -- (c21.west);
\draw [thick,->] (b21) -- (c2);
\draw [thick,->] (a21) -- (g2);
\draw [thick,->] (f2) -- (a2);
\node at (4.4,-1.6) [] (labb) {\normalsize (b)};

\end{tikzpicture}
\caption{Posets in Uniflow Partitions}
\label{fig:uniflowpart}
\end{figure}
\end{minipage}
\vspace{-0.3in}
\end{wrapfigure}
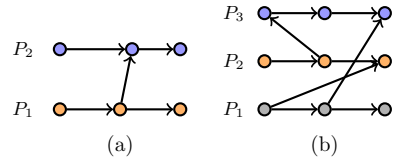

A uniflow partition of a computation's poset $P = (E,\ra)$ is its partition into \nuni chains $\{P_i ~|~ 1 \leq i \leq n_u\}$ 
such that no element (event of $E$) in a higher numbered chain is smaller than any element in lower numbered chain; that is if any event $e$ is placed 
on a chain $i$ then all causal dependencies of $e$ must be placed 
on chains numbered lower than $i$.  
For poset $P = (E,\ra)$, chain partition $\mu$ is uniflow if 
\begin{equation}
\label{eq:pos}
 \forall x,y \in P: \mu(x) < \mu(y) \Ra  \neg (y \not\ra x) 
 \end{equation}


\tikzstyle{place}=[circle,draw=black,fill=black,thick,inner sep=0pt,minimum size=2mm]
\tikzstyle{nsplace}=[circle,draw=black,thick,inner sep=0pt,minimum size=2mm]
\tikzstyle{place1}=[circle,draw=black,thick,inner sep=0pt,minimum size=3mm]
\tikzstyle{inner}=[circle,draw=black,fill=black,thick,inner sep=0pt,minimum size=1.25mm]
\tikzstyle{satlatnode}=[ellipse,draw=black,fill=black!20,thick,minimum size=5mm]
\tikzstyle{nsatlatnode}=[ellipse,style=dashed,thick]
\tikzstyle{satblock} = [rectangle, draw=gray, thin, fill=black!20,
text width=2.5em, text centered, rounded corners, minimum height=2em]
\tikzstyle{nsatblock} = [rectangle, draw=gray, thin,
text width=2.5em, text centered, rounded corners, minimum height=2em]
\tikzstyle{jbblock} = [rectangle, draw=black, thick, fill=black!20,
text width=2.5em, text centered, minimum height=2em]
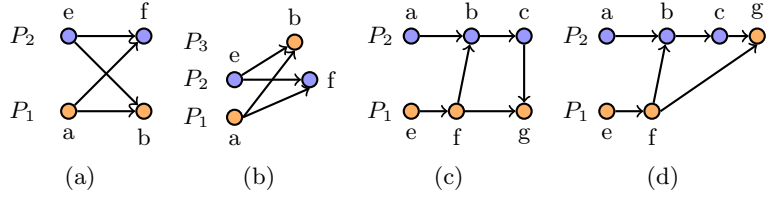
\begin{figure}[!t]
\centering
\subfloat[]{
\begin{tikzpicture}
\node at ( 0,1) [nsplace] [fill=blue!40] [label={above:{\small e}}]  (a) {};
\node at ( 1,1) [nsplace] [fill=blue!40] [label={above:{\small f}}] (b) {};
\node at ( 0,0) [nsplace] [fill=orange!60] [label={below:{\small a}}] (e) {};
\node at ( 1,0) [nsplace] [fill=orange!60] [label={below:{\small b}}] (f) {};
\draw [thick,->] (a.east) -- (b.west);
\draw [thick,->] (e.east) -- (f.west);
\draw [thick,->] (a) -- (f);
\draw [thick,->] (e) -- (b);
\node at (-0.2, 1) [label=left:$P_{2}$] (bot1) {};
\node at (-0.2, 0) [label=left:$P_{1}$] (bot2) {};
\remove{
\node at (4,-.15) [label=below:$(b)$] (subref1) {};
\node at ( 3.4,1) [nsplace] [fill=blue!40] (a2) {};
\node at ( 4.4,1) [nsplace] [fill=blue!40] (b2) {};
\node at ( 5.4,1) [nsplace] [fill=blue!40] (c2) {};
\node at ( 3.4,0) [nsplace] [fill=orange!60] (e2) {};
\node at ( 4.4,0) [nsplace] [fill=orange!60] (f2) {};
\node at ( 5.4,0) [nsplace] [fill=orange!60] (g2) {};
\draw [thick,->] (a2.east) -- (b2.west);
\draw [thick,->] (b2.east) -- (c2.west);
\draw [thick,->] (e2.east) -- (f2.west);
\draw [thick,->] (f2.east) -- (g2.west);
\node at (3.3, 1) [label=left:$P_{1}$] (bot1) {};
\node at (3.3, 0) [label=left:$P_{2}$] (bot2) {};
\node at (3.4,-1) [nsplace] [fill=green!60] (a21) {};
\node at (4.4,-1) [nsplace] [fill=green!60] (b21) {};
\node at (5.4,-1) [nsplace] [fill=green!60] (c21) {};
\draw [thick,->] (a21.east) -- (b21.west);
\draw [thick,->] (b21.east) -- (c21.west);
\node at (3.3, -1) [label=left:$P_{3}$] (bot21) {};
\draw [thick,->] (a21) -- (f2);
\draw [thick,->] (b2) -- (c21);
\node at (4,-1.15) [label=below:$(c)$] (subref3) {};
}
\end{tikzpicture}
}
\subfloat[]{
\begin{tikzpicture}
\node at ( 0,0.5) [nsplace] [fill=blue!40] [label={\small e}] (a) {};
\node at ( 1,.5) [nsplace] [fill=blue!40] [label={right:{\small f}}](b) {};
\node at ( 0,0) [nsplace] [fill=orange!60] [label={below:{\small a}}](e) {};
\node at ( 0.8,1) [nsplace] [fill=orange!60] [label={\small b}](f) {};
\draw [thick,->] (a.east) -- (b.west);
\draw [thick,->] (e.east) -- (f.south);
\draw [thick,->] (a) -- (f);
\draw [thick,->] (e.east) -- (b.south);
\node at (-0.1, 1) [label=left:$P_{3}$] (bot1) {};
\node at (-0.1, 0.5) [label=left:$P_{2}$] (bot2) {};
\node at (-0.1, 0) [label=left:$P_{1}$] (bot2) {};
\end{tikzpicture}
}
\subfloat[]{
\begin{tikzpicture}
\node at (0,1) [nsplace] [fill=blue!40] [label={above:{\small a}}] (a1) {};
\node at (0.8,1) [nsplace] [fill=blue!40] [label={above:{\small b}}](b1) {};
\node at (1.5,1) [nsplace] [fill=blue!40] [label={above:{\small c}}](c1) {};
\node at (0,0) [nsplace] [fill=orange!60] [label={below:{\small e}}](e1) {};
\node at (0.6,0) [nsplace] [fill=orange!60] [label={below:{\small f}}] (f1) {};
\node at (1.5,0) [nsplace] [fill=orange!60] [label={below:{\small g}}](g1) {};
\draw [thick,->] (a1.east) -- (b1.west);
\draw [thick,->] (b1.east) -- (c1.west);
\draw [thick,->] (e1.east) -- (f1.west);
\draw [thick,->] (f1.east) -- (g1.west);
\node at (0, 1) [label=left:$P_{2}$] (bot11) {};
\node at (0, 0) [label=left:$P_{1}$] (bot12) {};
\draw [thick,->] (f1) -- (b1);
\draw [thick,->] (c1) -- (g1);
\end{tikzpicture}
}
\subfloat[]{
\begin{tikzpicture}
\node at (0,1) [nsplace] [fill=blue!40] [label={above:{\small a}}](a1) {};
\node at (0.8,1) [nsplace] [fill=blue!40] [label={above:{\small b}}](b1) {};
\node at (1.5,1) [nsplace] [fill=blue!40][label={above:{\small c}}] (c1) {};
\node at (0,0) [nsplace] [fill=orange!60] [label={below:{\small e}}](e1) {};
\node at (0.6,0) [nsplace] [fill=orange!60] [label={below:{\small f}}] (f1) {};
\node at (2,1) [nsplace] [fill=orange!60] [label={above:{\small g}}](g1) {};
\draw [thick,->] (a1.east) -- (b1.west);
\draw [thick,->] (b1.east) -- (c1.west);
\draw [thick,->] (e1.east) -- (f1.west);
\draw [thick,->] (f1.east) -- (g1.south);
\node at (0, 1) [label=left:$P_{2}$] (bot11) {};
\node at (0, 0) [label=left:$P_{1}$] (bot12) {};
\draw [thick,->] (f1) -- (b1);
\draw [thick,->] (c1) -- (g1);
\end{tikzpicture}
}
\caption{Posets in (a) and (c) are not in uniflow partition: 
but (b) and (d) respectively are their equivalent uniflow partitions}
\label{fig:notuni}
\end{figure}


Visually, in a uniflow chain partition all the 
edges, capturing happened-before relation, between separate chains always point upwards 
because their dependencies --- elements of poset 
that are smaller --- are always placed on 
lower chains. 
Fig. \ref{fig:uniflowpart} shows two posets with uniflow partition. Whereas  
Fig. \ref{fig:notuni} shows two posets with 
partitions that do not satisfy the uniflow property. 
The poset in Fig.~\ref{fig:notuni}(a) can be 
transformed into a uniflow partition of three chains as shown in 
Fig. \ref{fig:notuni}(b). Similarly, Fig.~\ref{fig:notuni}(c) can be transformed into a uniflow partition 
of two chains shown in Fig. \ref{fig:notuni}(d).
Observe that: 

\begin{l1}
\label{lem:topo}
Every poset has at least one uniflow chain partition.
\end{l1}
\begin{proof}
Any total order derived from the poset is a uniflow chain partition in which 
 each element is a chain by itself.
In this trivial uniflow chain partition  
 the number 
of chains is equal to the number of elements in the 
poset. 
\remove{ Any non-trivial 
uniflow partition of the poset will place at least two elements 
on the same chain, and thus will have fewer 
chains than the size of the poset. Hence, for   
any poset $P$, the number of chains 
in any of its uniflow partition is always less 
than or equal to $|P|$.  }
\end{proof}

\remove{
 Generally, it is possible to obtain a 
uniflow partition with fewer chains. We discuss 
two  
 such approaches in Sec.~\ref{sec:partition}. 
 However, it is important to note that  
the number of chains in any uniflow partition 
of a poset is  always less than or equal to 
the size of the poset.

\begin{l1}
\label{lem:chainbound}
Let $P$ be a poset, and $\mu$ be its uniflow 
chain partition that partitions the poset 
in $n_u$ chains. Then, $n_u \leq |P|$.  
\end{l1}
\begin{proof}
Consider any trivial uniflow chain partition 
formed by using one total order derived from the poset. 
For each such uniflow partition  
 each element of the poset is a chain by itself, and 
 we have 
\end{proof}
}

The structure of uniflow chain partitions 
can be used for efficiently obtaining consistent 
cuts of larger ranks.   
\remove{
Given any consistent cut of the poset with 
uniflow chain partition, we can efficiently obtain
lexically     
bigger consistent cuts of larger ranks.}   
 
\begin{l1}[Uniflow Cuts Lemma]
\label{lem:uni}
Let $P$ be a poset with a
 uniflow chain partition
$\{P_i ~|~ 1 \leq i \leq n_u\}$, and $G$ be a 
consistent cut of $P$. Then 
any $H_k \subseteq P$ for $1 \leq k \leq n_u$ is also a consistent cut of $P$
if it satisfies:
\begin{center}
~~~~~~ $\forall i:k < i \leq n_u : H_k[i] = G[i]$, 
and\\
$\forall i:1 \leq i \leq k : H_k[i] = |P_i|$.
\end{center}
\end{l1}
\begin{proof}
Using  Equation \ref{eq:pos}, we 
 exploit the structure of uniflow 
chain partitions: the causal dependencies of 
any element $e$ lie only on   
chains 
that are lower than $e$'s chain. 
As $G$ is consistent, and $H_k$ contains the same 
elements as $G$ for the top $n_u - k$ chains, 
all the causal dependencies that 
need to be satisfied to make $H_k$ have to be on chain $k$ 
or lower. Hence, including all the 
elements from all of the lower chains will naturally 
satisfy all the causal dependencies, and make 
 $H_k$ consistent.   
\end{proof}

For example, in Fig. \ref{fig:uniflowpart}(b), consider the cut $G = [1,2,1]$ that is a consistent cut of the 
poset. Then, picking $k = 1$, and using Lemma \ref{lem:uni} gives us the cut 
$[1,2,3]$ which is consistent; similarly choosing 
$k = 2$ gives us $[1,3,3]$ that is also consistent. 
Note that the claim may not hold if the chain partition does not have uniflow property.
For example, in Fig. \ref{fig:notuni}(c), $G = [2,2]$ 
is a consistent cut.
The chain partition, however, is not uniflow and thus
applying the Lemma with $k = 1$ gives us $[2,3]$ 
which is not a consistent 
cut as it includes the third event on $P_1$, 
but not its causal dependency --- the third event 
on $P_2$.

\subsection{Finding a Uniflow Partition}

The problem of finding a uniflow chain partition is a direct extension 
of finding the {\em jump number} of a poset \cite{jump1,jump2,jump3}. 
Multiple algorithms have been proposed to find the jump number of a poset; 
which in turn arrange the poset in a uniflow chain partition. 
Finding an optimal (smallest number of chains) uniflow chain 
partition of a poset is a hard problem \cite{jump1,jump2}.  Bianco et al. \cite{jump2} present 
a heuristic algorithm to find a uniflow partition, and show in their 
experimental evaluation that in most of the cases the resulting partitions are   
relatively close to optimal.
We use a vector clock based online algorithm to find a 
uniflow partition for a computation. The details of this algorithm are given in 
Appendix~\ref{sec:partition}. Note that we need to re-generate vector clocks 
of the events for the uniflow partition. This is a simple task using 
existing vector clock implementation techniques, 
and we omit these details.  

\section{Polynomial Space Breadth-First Traversal of 
Consistent Cuts}
\label{sec:traverse}

\remove{Breadth-first traversal of lattice of consistent 
cuts provides many advantages
in analysis of parallel programs; its space requirement,
however, can be often prohibitive in analyzing large computations. }
BFS traversal of the lattice of consistent 
 cuts of any poset can be performed in space that is 
 polynomial in the size of the poset. We do so by first
 obtaining the poset's  
 uniflow chain partition, and then using this 
 partition for traversal of cuts in 
 increasing order of ranks.  
We start from the empty cut, and then traverse   all consistent cuts of rank $1$, then all consistent cuts of rank $2$ and so on.  For rank $r$, $1 \leq r \leq |E|$, we traverse the consistent cuts  
in the following lexical order: 
\vspace{1ex}
\begin{d1}[Lexical Order on Consistent Cuts]
Given any chain partition of poset $P$ that partitions it into $n$ chains,
we define a total order called {\em lexical order} on all consistent
cuts of $P$ as follows.
Let $G$ and $H$ be any two consistent cuts of $P$. Then,
$G <_l H \equiv \exists k: (G[k] < H[k]) \wedge (\forall i: n \geq i > k: G[i] = H[i])$ 
\end{d1}
\vspace{1.5ex}
 
\begin{wrapfigure}{R}{.46\textwidth}
\noindent
\begin{minipage}{.46\textwidth}
\vspace{-6ex}
\centering
\tikzstyle{place}=[circle,draw=black,fill=black,thick,inner sep=0pt,minimum size=2mm]
\tikzstyle{nsplace}=[circle,draw=black,thick,inner sep=0pt,minimum size=2mm]
\tikzstyle{place1}=[circle,draw=black,thick,inner sep=0pt,minimum size=3mm]
\tikzstyle{inner}=[circle,draw=black,fill=black,thick,inner sep=0pt,minimum size=1.25mm]
\tikzstyle{satlatnode}=[ellipse,draw=black,fill=black!20,thick,minimum size=5mm]
\tikzstyle{nsatlatnode}=[ellipse,style=dashed,thick]
\tikzstyle{satblock} = [rectangle, draw=gray, thin, fill=black!20,
text width=2.5em, text centered, rounded corners, minimum height=2em]
\tikzstyle{nsatblock} = [rectangle, draw=gray, thin,
text width=2.5em, text centered, rounded corners, minimum height=2em]
\tikzstyle{jbblock} = [rectangle, draw=black, thick, fill=black!20,
text width=2.5em, text centered, minimum height=2em]
\vspace{-6ex}
\begin{figure}[H]
\subfloat[]{
\begin{tikzpicture}
\node at ( 0,1) [nsplace] [fill=blue!40] [label={above:{\scriptsize $[1,0]$}}]  (a) {};
\node at ( 1,1) [nsplace] [fill=blue!40] [label={above:{\scriptsize $[2,1]$}}] (b) {};
\node at ( 0,0) [nsplace] [fill=orange!60] [label={below:{\scriptsize $[0,1]$}}] (e) {};
\node at ( 1,0) [nsplace] [fill=orange!60] [label={below:{\scriptsize $[1,2]$}}] (f) {};
\draw [thick,->] (a.east) -- (b.west);
\draw [thick,->] (e.east) -- (f.west);
\draw [thick,->] (a) -- (f);
\draw [thick,->] (e) -- (b);
\end{tikzpicture}
}
\qquad
\subfloat[]{
\begin{tikzpicture}
\node at ( 0,0.5) [nsplace] [fill=blue!40] [label={\scriptsize $[0,1,0]$}] (a) {};
\node at ( 1,.5) [nsplace] [fill=blue!40] [label={right:{\scriptsize $[0,2,1]$}}](b) {};
\node at ( 0,0) [nsplace] [fill=orange!60] [label={below:{\scriptsize $[0,0,1]$}}](e) {};
\node at ( 0.8,1) [nsplace] [fill=orange!60] [label={\scriptsize $[1,1,1]$}](f) {};
\draw [thick,->] (a.east) -- (b.west);
\draw [thick,->] (e.east) -- (f.south);
\draw [thick,->] (a) -- (f);
\draw [thick,->] (e.east) -- (b.south);
\end{tikzpicture}
}
\caption{Vector clocks of a computation in its original form, and in its uniflow partition}
\label{fig:transformvc}
\vspace{-2ex}
\end{figure}
\end{minipage}
\end{wrapfigure}
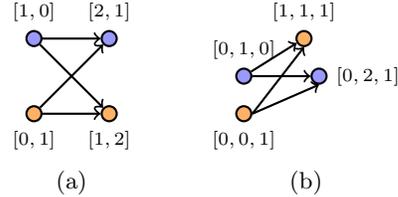
Recall from our vector clock notation (Section~\ref{sec:model}) that 
the right most entry in the vector clock is for the least significant (lowest) chain. Consider the poset 
with a non-uniflow chain partition in 
 \figref{fig:transformvc}(a). The vector clocks 
 of its events are shown against the four events. The  lexical order on the consistent cuts of this chain 
 partition is:
 $[0,0] <_l [0,1] <_l [1,0] <_l [1,1] <_l [1,2] <_l [2,1] <_l [2,2]$. 
For the same poset, \figref{fig:transformvc}(b)  shows the equivalent uniflow partition, and 
the corresponding vector clocks. The lexical order on the consistent cuts for this uniflow chain
partition is:
$[0,0,0] <_l [0,0,1] <_l [0,1,0] <_l [0,1,1] <_l [0,2,1] <_l [1,1,1] 
<_l [1,2,1]$. 

Note that the number of consistent cuts remains same 
for both of these chain partitions, and there is a 
one-to-one mapping between the consistent cuts 
in the two partitions. 
\begin{algorithm}[tbh]
\caption{{\sc TraverseBFSUniflow}($P$)}
\label{alg:outline}
\begin{algorithmic}[1]
\Require A poset $P = (E, \ra)$ that has been partitioned
into a uniflow chain partition of $n_u$ chains, and the
vector clock 
of the events have been regenerated for this partition.


\State {$G =$ new int[$n_u$]} \Comment{\textcolor{blue}{initial consistent cut}}
\State {enumerate($G$)} \Comment{\textcolor{blue}{evaluate the predicate on empty cut $G$.}}
\For{$(r = 1; r \leq |E|; r++)$}
\State{\textcolor{blue}{//make $G$ lexically smallest cut of given rank}}
\State {$G =$ \Call{GetMinCut}{$G,r$}} 

\While {$G \neq$ null}

\State {enumerate($G$)} \Comment{\textcolor{blue}{evaluate the predicate on $G$.}}
\State{\textcolor{blue}{//find 
the next bigger lexical cut of same rank}}
\State {$G =$ \Call{GetSuccessor}{$G,r$}} 
\EndWhile
\EndFor
\end{algorithmic}
\end{algorithm}

  Algorithm
 ~\ref{alg:outline}  shows the steps of our 
BFS traversal 
using a
 computation in a uniflow chain 
 partition. From Lemma~\ref{lem:topo}, we know that  
every poset has a uniflow chain partition. Recall 
 that the vector clocks of the events depend 
 on the chain partition of
the  poset. Thus, in generating 
this input we need 
two pre-processing steps: (a) finding a uniflow 
partition, and 
(b) regenerating vector clocks for this partition. 
For example, given a computation on two processes 
shown in \figref{fig:transformvc}(a), we will 
first convert it to the 
computation shown in \figref{fig:transformvc}(b). 
 These steps
are performed only once for a computation, and 
are relatively 
 inexpensive in comparison to the traversal 
 of lattice.  
\begin{wrapfigure}{R}{.5\textwidth}
\vspace{-8ex}
\begin{minipage}{.5\textwidth}
\input{algo/mincut}
\end{minipage}%
\vspace{-2ex}
\end{wrapfigure}
For each rank $r$, $1 \leq r \leq |E|$, Algorithm~\ref{alg:outline} first finds the lexically smallest consistent cut at of rank $r$.
This is done by the \getmin (shown in
Algorithm~\ref{alg:mincut}) routine that returns the lexically smallest 
consistent cut of $P$ bigger than $G$ of rank $r$. 
For example, in \figref{fig:illsone},
{\sc GetMinCut}($[0,0,0], 4$) returns $[0,1,3]$.
Given a consistent cut $G$ of rank $r$, we repeatedly find the next lexically bigger consistent cut of rank $r$
using the routine \getsucc given in Algorithm~\ref{alg:succ}.
For example, in \figref{fig:illsone},
{\sc GetSuccessor}$([0,0,3], 3)$ returns the next lexically smallest consistent cut $[0,1,2]$.

The \getmin routine on poset $P$ assumes that the rank of $G$ is at most $r$
and that $G$ is a consistent cut of the  $P$.
It first computes $d$ as the difference between $r$ and the rank of $G$.
We need to add $d$ elements to $G$ to find the smallest consistent cut of rank $r$.
We exploit the Uniflow Cut Lemma (Lemma \ref{lem:uni})
by adding as many elements from the lowest chain as possible.
If all the elements from the lowest chain are already in $G$, then we continue 
with the second lowest chain, and so on.
\begin{wrapfigure}{R}{0.52\textwidth}
\begin{minipage}{0.52\textwidth}
\vspace{-12ex}
\tikzstyle{place}=[circle,draw=black,fill=black,thick,inner sep=0pt,minimum size=2mm]
\tikzstyle{nsplace}=[circle,draw=black,thick,inner sep=0pt,minimum size=2mm]
\tikzstyle{place1}=[circle,draw=black,thick,inner sep=0pt,minimum size=3mm]
\tikzstyle{inner}=[circle,draw=black,fill=black,thick,inner sep=0pt,minimum size=1.25mm]
\tikzstyle{satlatnode}=[ellipse,draw=black,fill=black!20,thick,minimum size=5mm]
\tikzstyle{nsatlatnode}=[ellipse,style=dashed,thick]
\tikzstyle{satblock} = [rectangle, draw=gray, thin, fill=black!20,
text width=2.5em, text centered, rounded corners, minimum height=2em]
\tikzstyle{nsatblock} = [rectangle, draw=gray, thin,
text width=2.5em, text centered, rounded corners, minimum height=2em]
\tikzstyle{jbblock} = [rectangle, draw=black, thick, fill=black!20,
text width=2.5em, text centered, minimum height=2em]
\begin{figure}[H]
\centering
\begin{tikzpicture}
\node at (-0.6,0.8) []  (t3) {$P_3$};
\node at (0,0.8) [nsplace] [fill=blue!40][label={above:{\scriptsize $[1,2,0]$}}]  (a2) {};
\node at (1,0.8) [nsplace] [fill=blue!40] [label={above:{\scriptsize $[2,2,0]$}}] (b2) {};
\node at (2,0.8) [nsplace] [fill=blue!40] [label={above:{\scriptsize $[3,2,2]$}}] (c2) {};
\node at (-.6,0) []  (t3) {$P_2$};
\node at (0,0) [nsplace] [fill=orange!60] [label={above:{\scriptsize $[0,1,0]$}}](e2) {};
\node at (1,0) [nsplace] [fill=orange!60] [label={above:{\scriptsize $[0,2,0]$}}](f2) {};
\node at (2,0) [nsplace] [fill=orange!60] [label={above:{\scriptsize $[0,3,1]$}}](g2) {};
\draw [thick,->] (a2.east) -- (b2.west);
\draw [thick,->] (b2.east) -- (c2.west);
\draw [thick,->] (e2.east) -- (f2.west);
\draw [thick,->] (f2.east) -- (g2.west);
\node at (-.6,-.8) []  (t3) {$P_1$};
\node at (0,-0.8) [nsplace] [fill=gray!60] [label={below:{\scriptsize $[0,0,1]$}}] (a21) {};
\node at (1,-0.8) [nsplace] [fill=gray!60] [label={below:{\scriptsize $[0,0,2]$}}](b21) {};
\node at (2,-0.8) [nsplace] [fill=gray!60] [label={below:{\scriptsize $[0,0,3]$}}](c21) {};
\draw [thick,->] (a21.east) -- (b21.west);
\draw [thick,->] (b21.east) -- (c21.west);
\draw [thick,->] (b21) -- (c2);
\draw [thick,->] (a21) -- (g2);
\draw [thick,->] (f2) -- (a2);
\end{tikzpicture}
\caption{Illustration: \getsucc}
\label{fig:illsone}
\end{figure}
\vspace{-7ex}
\input{algo/getsucc}
\vspace{-10ex}
\end{minipage}
\end{wrapfigure}
For example in \figref{fig:illsone},
consider finding smallest consistent cut of rank $5$ starting from $G=[0,0,2]$.
In this case, we add all three elements from $P_1$ to reach
 $[0,0,3]$, and then add first two elements from $P_2$ to get the answer as $[0,2,3]$.

The \getsucc routine (Algorithm~\ref{alg:succ})  
 finds the lexical successor of $G$ at  rank $r$. The approach for finding a lexical successor is similar to 
 counting numbers in a decimal system: if we are looking for successor of 2199, then we can't increment 
 the two 9s (as we are only allowed digits 0-9), and hence the first possible increment is for  
entry 1. We increment it to 2, but we must now reset the entries at lesser significant digits. 
 Hence, we reset the two 9s to 0s, and get the successor as 2200. 

In our \getsucc routine, 
we start at the second lowest 
 chain in a uniflow poset, and if possible increment the cut by one event on 
 this chain. We then reset the entries on   
 lower chains, and then make the cut consistent by satisfying all the causal dependencies. 
 If the rank of the resulting cut is less than or equal to $r$, then  
calling the \getmin routine gives us the lexical successor of $G$ at rank $r$. 
 
 Line 1 copies cut $G$ in $K$.
The {\sf for} loop covering lines 2--13 searches for 
an appropriate element not in $G$ such that adding this element makes
the resulting consistent cut lexically greater than $G$.
We start the search from chain $2$, 
instead of chain $1$, because
 for a non-empty cut $G$ adding any event from the lowest chain to $G$ will only increase $G$'s rank as there are no lower chains to reset.
Line 3 checks if there is any possible element to add in $P_i$. If yes, then 
\remove{, if not
If all elements from $P_i$ are already included in $G$, we go to the next value of $i$.}
lines 4--6 increment $K$ at chain $i$, and then 
set all its values for lower chains to $0$.
To ensure that $K$ is a consistent cut, for every element in $K$, we add its causal dependencies 
to $K$ in lines 7--11. Line 12 checks whether the resulting consistent cut 
is of rank $\leq r$. If $rank(K)$ is at most $r$, then we have found a suitable cut that
can be used to find the next lexically bigger consistent cut and we call \getmin
routine to find it. 
If we have tried all values of $i$ and did not find a suitable cut, then 
$G$ is the largest consistent cut of rank $r$ and we return
{\sf null}.

In \figref{fig:illsone}, consider the call of 
\getsucc($[1,2,3],6$). As there is no next element in $P_1$, we consider the next element
in $P_2$. After line 5, the value of $K$ is $[1,3,0]$, 
which is not consistent. Lines 7--10 
make $K$ a consistent cut, now $K = [1,3,1]$. 
Since $rank(K)$ is $5$, we call 
{\sc GetMinCut} at line 13 to find the smallest consistent cut of rank $6$ that is lexically bigger than $[1,3,1]$.
This consistent cut is $[1,3,2]$.

We present the proof of correctness of in Appendix~\ref{app:proofs}. 

\subsection{Optimization for Time Complexity}  
\label{sec:opttime}

We can find the lexical successor of any 
consistent cut 
in $\bigo{n_u^2}$ time, instead of $\bigo{n_u^3}$ time taken in 
\getsucc, by using additional $\bigo{n_u^2}$ space. 

\begin{wrapfigure}{R}{.6\textwidth}
\vspace{-7ex}
\begin{minipage}{.6\textwidth}
\begin{algorithm}[H]
\caption{{\sc ComputeProjections}($G$)}
\label{alg:joins}
\begin{algorithmic}[1]
\Require $G$: a consistent cut of rank $r$

\For{($i = n_u; i \geq 1; i--$)} \Comment{\textcolor{blue}{go top to bottom}}
\State {$val = G[i]$} \Comment{\textcolor{blue}{event number in $G$ on chain $i$}}

   \State {$vc = $ vector clock of event num $val$ on chain $i$}
   \If{$i == n_u$} \Comment{\textcolor{blue}{on highest chain}}
   \State {$proj[i] = vc$}
   \Else
\Comment{\textcolor{blue}{process relevant entries in vector}}
   \For{$(j = i; j > 0; j--)$} 
\State{\textcolor{blue}{//projection on chain $i$:}}
   \State {$proj[i][j] = $MAX$(vc[j],proj[i+1][j])$} 
   \EndFor
   \EndIf
\EndFor
\end{algorithmic}
\end{algorithm}
\end{minipage}%
\vspace{-3ex}
\end{wrapfigure}

Observe that \getsucc routine iterates over $n_u - 1$ chains in the outer loop
at line 2, and the two inner loops at lines 8 and 9 perform $\bigo{n_u^2}$ 
work in the worst case. When we cannot find a suitable cut of rank less than or equal to $r$ (check performed at line 12), we 
 move to a higher chain (with the outer loop at line 2). Thus, we repeat 
a large fraction of the $\bigo{n_u^2}$ work in the two inner loops 
at lines 8 and 9 for this higher chain. We can avoid this repetition by storing the combined 
causal dependencies from higher chains on each lower chain.
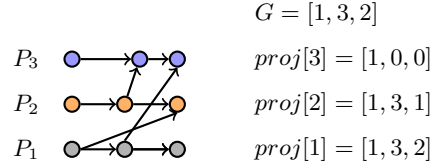
\begin{wrapfigure}{R}{0.47\textwidth}
\vspace{-8ex}
\begin{minipage}{0.47\textwidth}
\tikzstyle{place}=[circle,draw=black,fill=black,thick,inner sep=0pt,minimum size=2mm]
\tikzstyle{nsplace}=[circle,draw=black,thick,inner sep=0pt,minimum size=2mm]
\tikzstyle{place1}=[circle,draw=black,thick,inner sep=0pt,minimum size=3mm]
\tikzstyle{inner}=[circle,draw=black,fill=black,thick,inner sep=0pt,minimum size=1.25mm]
\tikzstyle{satlatnode}=[ellipse,draw=black,fill=black!20,thick,minimum size=5mm]
\tikzstyle{nsatlatnode}=[ellipse,style=dashed,thick]
\tikzstyle{satblock} = [rectangle, draw=gray, thin, fill=black!20,
text width=2.5em, text centered, rounded corners, minimum height=2em]
\tikzstyle{nsatblock} = [rectangle, draw=gray, thin,
text width=2.5em, text centered, rounded corners, minimum height=2em]
\tikzstyle{jbblock} = [rectangle, draw=black, thick, fill=black!20,
text width=2.5em, text centered, minimum height=2em]
\begin{figure}[H]
\centering
\begin{tikzpicture}
\node at ( 0,1.2) [nsplace] [fill=blue!40] (a) {};
\node at ( 0.9,1.2) [nsplace] [fill=blue!40] (b) {};
\node at ( 1.4,1.2) [nsplace] [fill=blue!40] (c) {};
\node at ( 0,0.6) [nsplace] [fill=orange!60] (e) {};
\node at ( 0.7,0.6) [nsplace] [fill=orange!60] (f) {};
\node at ( 1.4,.6) [nsplace] [fill=orange!60] (g) {};

\node at (0,0) [nsplace] [fill=gray!60] (a21) {};
\node at (0.7,0) [nsplace] [fill=gray!60] (b21) {};
\node at (1.4,0) [nsplace] [fill=gray!60] (c21) {};

\draw [thick,->] (a.east) -- (b.west);
\draw [thick,->] (b.east) -- (c.west);
\draw [thick,->] (e.east) -- (f.west);
\draw [thick,->] (f.east) -- (g.west);
\draw [thick,->] (f) -- (b);
\node at (-0.2, 0) [label=left:$P_{1}$] (bot1) {};
\node at (-0.2, 0.6) [label=left:$P_{2}$] (bot2) {};
\node at (-0.2, 1.2) [label=left:$P_{3}$] (bot3) {};

\draw [thick,->] (a21.east) -- (b21.west);
\draw [thick,->] (b21.east) -- (c21.west);
\draw [thick,->] (b21.north) -- (c.south);
\draw [thick,->] (b21.east) -- (c21.west);
\draw [thick,->] (a21.east) -- (g.south);

\node at (2.2, 1.8) [label=right:{$G=[1,3,2]$}] (bot1) {};
\node at (2.2, 1.2) [label=right:{$proj[3]=[1,0,0]$}] (bot1) {};
\node at (2.2, 0.6) [label=right:{$proj[2]=[1,3,1]$}] (bot2) {};
\node at (2.2, 0) [label=right:{$proj[1]=[1,3,2]$}] (bot3) {};
\end{tikzpicture}
\caption{Projections of a cut on chains}
\label{fig:joins}
\end{figure}
\end{minipage}
\vspace{-2ex}
\end{wrapfigure}

Let us illustrate this with an example. Consider 
the uniflow computation shown in \figref{fig:joins}. 
Suppose we want the lexical successor
of $G = [1,3,2]$. Then, 
for each chain, starting from 
the  top we compute 
the projection of events included in $G$ on lower chains.  
For example, $G[3] = 1$, and thus on the top-most chain, 
the  projection is only the 
vector clock of the first event on $P_3$, which is $[1,0,0]$. Thus $proj[3] = [1,0,0]$. 
On $P_2$, the projection must include the 
combined vector clocks of $G[3]$ and $G[2]$ ---
the events from top two chains. As $G[2] = 3$, 
we use the vector clock of third event on $P_2$, which 
is 
 $[0,3,1]$ as that event 
 is causally dependent on first event on $P_1$.
Combining the two vectors gives us 
the projection on $P_2$ as $proj[2] = [1,3,1]$. 
\remove{
Note that combined projection on the lowest chain 
in a uniflow chain partitioned poset  
will always be exactly the same as $G$ (and thus for our implementation, we can skip computing it and can just 
copy its value as the vector clock representation 
of $G$). }

Algorithm~\ref{alg:joins} shows the steps 
involved in computing the projections of a cut 
on each chain. We create an auxiliary matrix, $proj$,
of size $n_u \times n_u$, 
to store these projections. 
In \getsucc routine, once we have computed a new successor 
by using some event on chain $i$, we need to update the 
stored projections on chains lower than $i$; 
and not all $n_u$ chains. This is because 
the projections for unchanged entries in $G$ above  
chain $i$ will not change on chain $i$, or any chain above it. Hence,  
we only update
the relevant rows and columns --- rows and columns 
with number $i$ 
or lower --- in $proj$; i.e. only the upper 
triangular part of the matrix $proj$. We keep track 
of the chain that gave us the successor cut, and pass it  
as an additional argument to Algorithm~\ref{alg:joins}. 
We   
read and update $n_u^2/2$ entries in the matrix, 
and not all $n_u^2$ of them. 
\remove{ Even though the Hence, using this approach 
we reduce the work within the outer loop (that 
starts at line 2) to $\bigo{n_u}$, and 
overall work per consistent cut to 
$\bigo{n_u^2}$.} 

Hence, the optimized implementation of finding the lexical 
successor of $G$ requires two changes. First, every call of \getsucc$(G,r)$ starts with 
first computing the projections of $G$ using \algoref{alg:joins}. 
Second, we replace the 
two inner {\sf for} loops at lines 8 and 9 in 
\getsucc by one $\bigo{n_u}$ loop to compute the 
max of the two vector clocks: vector clock of 
$K[i]$, and $proj[i]$.   
In interest of space, we show the optimized routine for \getsucc with 
these changes in Algorithm~\ref{alg:succopt} in Appendix~\ref{app:getsuccopt}. 
 
\subsection{Re-mapping Consistent Cuts to Original
Chain Partition}

The number 
of consistent cuts of a computation is independent 
 of the chain partition used. Their vector clock 
 representation, however, varies with chain 
 partitions as the vector clocks of events 
 in the computation depend on the chain partition 
 used to compute them. 
There is a one-to-one mapping between a consistent 
cut in the original chain partition of the computation 
on $n$ chains (processes), and its uniflow chain partition
on $n_u$ chains. We now show how to map a consistent cut in a uniflow chain
partition to its equivalent cut in the original chain partition 
of the computation. Let $P=(E,\ra)$ be a computation on $n$ 
processes, and let $n_u$ be the number of chains 
in its uniflow chain partition. If $G_u$ is a consistent 
cut in the uniflow chain partition, then its equivalent 
consistent cut 
$G$ for the original chain partition (of $n$ chains) can 
be found in $\bigo{n_u + n^2}$ time.  

We do so by mapping two additional entries 
with the new vector clock of each event for uniflow chain 
partition: the chain number $c$, and event number $e$
from the original chain partition over $n$ chains. 
For example, in \figref{fig:transformvc}(b), 
for uniflow vector clock $[1,1,1]$, its chain number
in original poset is $1$, and its event 
number on that chain is $2$. When generating 
the uniflow vector clocks, we populate these  
entries in a map. Given a uniflow vector clock 
$uvc$, the call to {\sc OriginalChain}$(uvc)$ returns 
$c$, and {\sc OriginalEvent}$(uvc)$ returns $e$. 
To compute $G$ from $G_u$, we use these 
two values from the corresponding event for each 
entry in $G_u$. We  
 start with $I$ as an all-zero vector of length $n$. Now,
 we iterate over $G_u$, and  
we update $I$ by setting $I[c] = max(I[c],
e)$. 
\begin{algorithm}[tbh]
\caption{{\sc Remap}($G_u, n_u, n$)}
\label{alg:remap}
\begin{algorithmic}[1]
\Require $G_u$: a consistent cut in uniflow chain partition on $n_u$ chains
\Ensure $G$: equivalent consistent cut in original
chain partition on $n$ chains

\State{$G = $ new int[$n$]} \Comment{\textcolor{blue}{allocate memory for $G$}}
\State{$I = $ new int[$n_u$]} \Comment{\textcolor{blue}{reduction vector}}

\For{($i = n_u; i \geq 1; i--$)} \Comment{\textcolor{blue}{go over all the uniflow chains}}
   \State {$uvc =$event number $G_u[i]$'s vector-clock on uniflow chain $i$}
\State{\textcolor{blue}{//chain of this event in original poset}}
   \State{$c =  \Call{OriginalChain}{uvc}$} 
\State{\textcolor{blue}{//$uvc$'s event number on chain $c$ in original poset}}
   \State{$e =  \Call{OriginalEvent}{uvc}$}    
   \If{$I[c] < e$} \Comment{\textcolor{blue}{update indicator with $e$}}
   \State {$I[c] = e$}
   \EndIf
\EndFor
\For{($j = n; i \geq 1; i--$)} \Comment{\textcolor{blue}{go over chains in original poset}}
\State {$vce = $event number $I[j]$'s vector-clock on chain $j$ in original poset}
\For{($k = n; k \geq 1; k--$)} \Comment{\textcolor{blue}{update $G$ entries}}
\State {$G[k] = $MAX$(G[k],vce[k])$} 
\EndFor
\EndFor

\State {\Return $G$}
\end{algorithmic}
\end{algorithm}

 As vector $G_u$ has length $n_u$, 
this step takes $\bigo{n_u}$ time.
We now initiate $G$ as an all-zero vector clock of 
length $n$, and 
 for each entry $I[k]$, $1 \leq k \leq n$, we get the  
vector clock, $vce$, of event $I[k]$ on chain $k$ in the 
original computation. We then set $G$ to the 
component-wise maximum of $G$ and $vce$. 
As there are $n$ entries in $I$, and for each non-zero
entry we perform $\bigo{n}$ work in updating $G$ (in lines 
11--14 in Algorithm~\ref{alg:remap})
the total work in this step is $\bigo{n^2}$. 

\subsection{Traversing consistent cuts of a given rank}

A key benefit of our algorithm is that it can 
traverse all the consistent cuts of a given rank, 
or within a range of ranks, without traversing the 
cuts of lower ranks. In contrast, 
the traditional BFS traversal must traverse, and 
store, consistent cuts of rank $R-1$ to traverse cuts 
of rank $R$, which in turn requires it to traverse cuts of rank $R-2$ and so on. 

To traverse all the cuts of rank $R$, we only need to change the loop 
bounds at line 3 in Algorithm~\ref{alg:outline} to 
{\sf for $(r=R; r \leq R;r++)$}. Thus, starting with an empty cut we can 
find the lexically smallest consistent cut of rank $r$ in $\bigo{n_u}$ time 
with the \getmin routine. Then we repeatedly find its lexical successor of the same 
rank, until we have traversed the lexically biggest cut of rank $R$. 
Similarly, consistent cuts between the ranks of $R_1$ and $R_2$ 
can be traversed by changing the loop 
 at line 3 in Algorithm~\ref{alg:outline} to: 
{\sf for $(r=R_1; r \leq R_2;r++)$}.
\medskip
\begin{l1}
Let $L_k$ denote the number of consistent cuts of rank $k$ for 
a computation $(E,\ra)$. Then, 
traversing  consistent cuts of rank $r$ takes $\bigo{n_u^2L_r}$ time with
Algorithm~\ref{alg:outline}. For the same traversal, the traditional BFS algorithm 
requires $\bigo{n^2\sum_{k=1}^{r} L_k}$ time, and Lex algorithm 
takes $\bigo{n^2\sum_{k=1}^{|E|} L_k}$ time.
\end{l1}

\section{Time and Space Complexity}

Algorithm~\ref{alg:outline} requires a computation in its 
uniflow chain partition. Multiple polynomial time algorithms 
exist to find a  non-trivial uniflow chain partition of a poset, and we give a vector clock 
based online algorithm to find 
one in Appendix~{sec:partition} that takes $\bigo{n}$ time per event. 
We analyze the worst case time and space complexities of our algorithms. 

  Given any 
 computation on $n$ processes and $E$ events, 
we can find   
its trivial uniflow chain partition in $\bigo{n|E|log|E|}$
time by lexically ordering the vector clocks of 
all the events. 
 
Suppose the number of chains in the uniflow partition 
is $n_u$, then
\remove{
In Section~\ref{sec:partition}, we present two algorithms 
to find non-trivial uniflow chain partitions.  }
 the step of computing 
new vector clocks takes $\bigo{n_u|E|\cdot \Delta}$ time where 
$\Delta$ is the maximum in-degree of  
any event in the computation; note that  $\Delta \leq n$. 
The \getmin sub-routine has only one {\sf for} loop 
that iterates over the chains of the uniflow partition. 
Hence, it takes $\bigo{n_u}$ time in the worst case. 
The optimized version of finding the successor, 
 sub-routine  \getsuccopt, takes $\bigo{n_u^2}$ time in 
the worst case due to the two nested
{\sf for} loops at lines 3, and 10. Hence, for any 
rank, our algorithm 
requires $\bigo{n_u^2}$ time per consistent cut 
in the uniflow partition. Re-mapping  this cut 
to the original computation takes 
$\bigo{n_u + n^2}$ time. Thus,  
we take $\bigo{n_u^2 + n^2}$ time per consistent 
cut.  
\medskip
\begin{t1}
Given a computation $P = (E,\ra)$ on $n$ processes, 
Algorithm~\ref{alg:outline} performs 
breadth-first traversal of its lattice
of consistent cuts using $\bigo{(n_u + n) |E|}$ space 
which is polynomial 
in the size of the computation. 
\end{t1}

\begin{proof}
Storing  the original computation  
requires $\bigo{n|E|}$ 
space --- each event's vector clock having at most $n$ integers. 
 Vector clocks for the 
uniflow chain partition with $n_u$ chains takes 
$\bigo{n_u}$ space per event. Thus, we require 
$\bigo{n_u} |E|$ additional space overall to store 
the computation in its uniflow form. 
Traversing the lattice as per Algorithm~\ref{alg:outline} 
only requires $\bigo{n_u^2}$ space as at most two vectors 
of length $n_u$ are stored/created during this traversal, and 
we use the auxiliary matrix of $n_u \times n_u$ size 
in the optimized implementation of \getsucc. 
\remove{
Our analysis above showed that 
the breadth-first traversal on the 
lattice of consistent cuts requires  $\bigo{(n_u|E|}$ space 
overall ---  $\bigo{n_u|E|}$ to store the vector clocks 
as per the uniflow chain partition, and  $\bigo{n_u}$ 
for traversing the lattice.} From Lemma~\ref{lem:topo} we know that 
$n_u \leq |E|$. Thus, the worst case space complexity of  
 is $\bigo{|E|^2 + n|E|}$ which is 
polynomial in the size of the input.   
\end{proof}

\newcommand{\our}{{\sf UniBFS}\xspace}
\newcommand{\bfs}{{\sf TBFS}\xspace}
\section{Experimental Evaluation}
\label{sec:eval}

We conduct an experimental evaluation to compare the 
space and time required by BFS, Lex, and our uniflow based traversal algorithm to traverse consistent cuts of specific ranks, as well 
as all consistent cuts up to a given rank. We do not evaluate DFS implementation as 
previous studies have shown that Lex implementation outperforms DFS based 
traversals in both time and space \cite{cite:enumeratingglobal,chang2015parallel,quicklex}. Lexical enumeration is significantly better for enumerating all 
possible consistent cuts of a computation \cite{chang2015parallel,quicklex}.
However, it is not well suited for only traversing cuts of a specified ranks, 
or finding the smallest counter example.  For these tasks, BFS traversal 
remains the algorithm of choice. 
We optimize the traditional BFS implementation as per \cite{cite:enumeratingglobal} to enumerate every global state exactly once.
We use seven benchmark computations from recent literature on 
traversal of consistent cuts \cite{chang2015parallel,quicklex}.   
 The details of these benchmarks are shown in first four columns of Table~\ref{tab:bench}. 
Benchmarks 
{\em d-100, d-300} and {\em d-500}
are randomly generated posets for modeling distributed computations.
The benchmarks {\em bank}, and {\em hedc}
are computations obtained from real-world concurrent programs that are used by \cite{cite:jpredictor,cite:fasttrack,cite:datarace} for evaluating their predicate detection algorithms. The benchmark {\em bank} contains a typical error pattern in concurrent programs, and  {\em hedc} is a web-crawler. 
 Benchmarks {\em w-4} and {\em w-8} have 480 events distributed over 4 and 8 processes 
respectively, and  
help to highlight the influence of degree of parallelism on the performance of enumeration algorithms.
\begin{table*}[]
\centering
\begin{tabular*}{\textwidth}{@{\extracolsep{\fill}}@{}lrrrrrrrrr@{}}
\toprule
\multicolumn{1}{l}{Name} & \multicolumn{1}{c}{$n$}                    & 
\multicolumn{1}{c}{$|E|$}
& \begin{tabular}[c]{@{}c@{}}Approx.\\ \# of cuts\end{tabular} & $n_u$                  & \begin{tabular}[c]{@{}c@{}}$T_{part}$\end{tabular}~~ & \multicolumn{2}{c}{\sf Traditional BFS}  & \multicolumn{2}{c}{\sf Uniflow BFS}  \\
                &                         &                           &                                                                                  &                           &                                & \multicolumn{1}{c}{~~Space}                  & \multicolumn{1}{c}{Time}                   & \multicolumn{1}{c}{~~Space}                & \multicolumn{1}{c}{Time} \\
\midrule
d-100 & 10 & 100 & 1.2$\times10^6$ & 26 & 0.030 & 108 & 0.48 & 31 & 0.37\\
d-300 & 10 & 300 & 4.3$\times10^7$ & 68 & 0.031 & 842 & 16.84 & 33 & 46.20\\
d-500 & 10 & 500 & 4.9$\times10^9$ & 112 &  0.033 & 893 & 108.07 & 34 & 607.55\\
bank & 8 & 96 & 8.2$\times10^8$ &  8 & 0.023 & \textcolor{red}{$\times$} & \textcolor{red}{$\times$} & 59 & 73.2\\
hedc & 12 & 216 & 4.5$\times10^9$ & 26  & 0.028 & \textcolor{red}{$\times$} & \textcolor{red}{$\times$} & 56 & 1129  \\
w-4 & 4 & 480 & 9.3$\times10^6$ &  121 & 0.036 & 258  & 0.99 & 25 & 8.59 \\
w-8 & 8 & 480 & 7.3$\times10^9$ & 63 & 0.032 & \textcolor{red}{$\times$} & \textcolor{red}{$\times$} & 40  & 1445.57\\
\bottomrule
\end{tabular*}
\caption{Benchmark details, heap-space consumed (in MB) and runtimes (in seconds) for two BFS implementations to traverse
the full lattice of consistent cuts. $T_{part}$= time (seconds) to find uniflow partition; \textcolor{red}{$\times$} = out-of-memory error}
\label{tab:bench}
\end{table*}

\begin{table}[!htb]
\centering
\caption{Runtimes (in seconds) for {\sf tbfs}: Traditional BFS, {\sf lex}: Lexical, and {\sf uni}: Uniflow BFS implementations to traverse cuts of given ranks}
\label{tab:times}
\begin{tabular*}{\textwidth}{@{\extracolsep{\fill}}l|rrr|rrr|rrr|rrr@{}}
\toprule
\multicolumn{1}{c}{Name} & \multicolumn{3}{c}{$r =\frac{|E|}{4}$}                                          & \multicolumn{3}{c}{$r =\frac{|E|}{2}$} & \multicolumn{3}{c}{$r =\frac{3|E|}{4}$} & \multicolumn{3}{c}{$r \leq 32$} \\ \midrule
                         & {\sf tbfs}                    & {\sf lex}                    & {\sf uni}                    & {\sf tbfs}       & {\sf lex}      & {\sf uni}      & {\sf tbfs}       & {\sf lex}       & {\sf uni} & {\sf tbfs}       & {\sf lex}       & {\sf uni}      \\ 
d-100    & 0.12   & 0.10   & 0.04  & 0.22 & 0.11   & 0.05    & 0.20  & 0.89   & 0.04  & 0.19   & 0.93   & 0.12  \\
d-300    & 0.39   & 1.23   & 0.05  & 2.70 & 1.15   & 0.07    & 6.33  & 1.25   & 0.13  & 0.20   & 1.22   & 0.14  \\
d-500    & 2.29  & 5.73   & 0.11 & 7.83 & 6.52   & 0.33   & 67.59 & 6.86   & 1.48 & 0.19   & 4.93   & 0.19  \\
bank     & 3.36  & 16.80  & 0.27 &\textcolor{red}{{$\times$}}  & 16.34  & 3.07  &\textcolor{red}{{$\times$}}   & 17.02  & 0.32  & 45.43 & 16.87  & 5.70\\
hedc     & 4.72  & 16.50 & 0.40 & \textcolor{red}{{$\times$}}  & 152.76 & 15.70 & \textcolor{red}{{$\times$}}   & 153.54 & 0.51  & 0.23   & 128.60 & 0.12   \\
w-4      & 0.09    & 0.18    & 0.07  & 0.53  & 0.18    & 0.10   & 0.93   & 0.19    & 0.09   & 0.01     & 0.13    & 0.05   \\
w-8      & 26.39 & 143.08 & 0.72 & \textcolor{red}{{$\times$}}  & 171.23 & 120.27 & \textcolor{red}{{$\times$}}   & 169.21 & 3.09 & 0.02    & 196.21 & 0.05   \\
\bottomrule
\end{tabular*}
\end{table}

We conduct two sets of experiments: (a) complete traversal of lattice of 
consistent cuts (of the computation) in BFS manner, and (b) traversal of 
 cuts of specific ranks.
We conduct all the experiments on a Linux machine with an Intel Core i7 3.4GHz CPU, with L1, L2 and L3 
caches of size 32KB, 256KB, and 8192KB respectively. We compile and run the programs  
on Oracle Java 1.7, and  
limit the maximum heap size for Java virtual machine (JVM) to 2GB. 
\remove{We call our implementation of 
uniflow chain partition based BFS  \our, and denote the traditional BFS implementation by \bfs. }
For each 
 run of our traversal algorithm, 
we use Algorithm~\ref{alg:onlinepart} (in Appendix~\ref{sec:partition}) to find 
 the uniflow chain partition of the poset. 
The runtimes and space reported for our uniflow traversal implementation include
the time and space
needed for finding and storing the uniflow chain partition 
of the poset. 

Table~\ref{tab:bench} compares the size of JVM heap and runtimes for traditional BFS 
and our uniflow based BFS  
 traversal of lattice of consistent cuts of the benchmarks.  The
traditional BFS implementations runs out of memory on {\em hedc, bank}, and {\em w-8}. Our 
implementation requires significantly less memory, and 
even though it is 
slower, it enables 
us to do BFS traversal on large computations --- something that is impossible  
with traditional BFS due to its memory requirement.  

Table~\ref{tab:times} highlights the strength of our algorithm 
in traversing consistent cuts of specific ranks. We compare our implementation 
with traditional BFS as well as the implementation of Lexical traversal. 
For traversing consistent cuts of three specified ranks (equal to quarter, half, and three-quarter
of number of events) our algorithm is consistently and significantly faster than both 
traditional BFS, as well as Lex algorithm. Thus, it can be extremely helpful 
in quickly analyzing traces when the programmer has knowledge of the conditions when 
an error/bug occurs. 

In addition, there are many cases when we are not interested in checking 
all consistent cuts of a computation.
It has been 
argued that most concurrency 
related bugs can be found relatively early in execution traces 
\cite{cite:chess,ball2010preemption}.
We also perform well in visiting  
all consistent cuts of rank less than or equal to 32. Hence, our implementation
is faster on most benchmarks for smaller ranks, and requires much 
less memory (memory consumption details for this experiment are in 
Appendix~\ref{app:memory32}).    
These results emphasize that our algorithm is useful for practical 
debugging tasks while consuming less resources.   

\remove{
 Each of them simulates a 
distributed computation on $n=10$ processes, with varying number of events: d-50 have 50 events, 
d-100 has 100, and d-300 has 300 events. After each internal event, every processes sends a message to a 
randomly selected process with a probability of 0.3.  
}

\remove{
The runtime values of the two implementations are shown the last two columns of Table~\ref{tab:bench}. 
Note that for four benchmarks that generate large number of consistent cuts, \bfs (traditional BFS) 
runs out of the 2GB memory allocated to JVM. For the remaining three benchmarks, \our is slightly faster  
on smaller posets, and is within the same order of magnitude for a large poset. This is because large posets
generally lead to larger number of chains in uniflow partition, and our algorithm requires $\bigo{n_u^2}$ time 
per consistent cut, whereas \bfs takes $\bigo{n^2}$ time. 
Given that \bfs needs exponential space for the traversal, and cannot traverse posets with large lattices,  \our is extremely useful for such cases.  
The time taken in finding the uniflow chain partitions, and regenerating the 
vector clocks of the events is show in Figure~\ref{fig:vcgen}. As this result shows, 
we do not spend too much time in the pre-processing steps and most of the time is consumed 
by the traversal of the lattice. 

\begin{figure}
\centering
\includegraphics[scale=0.5]{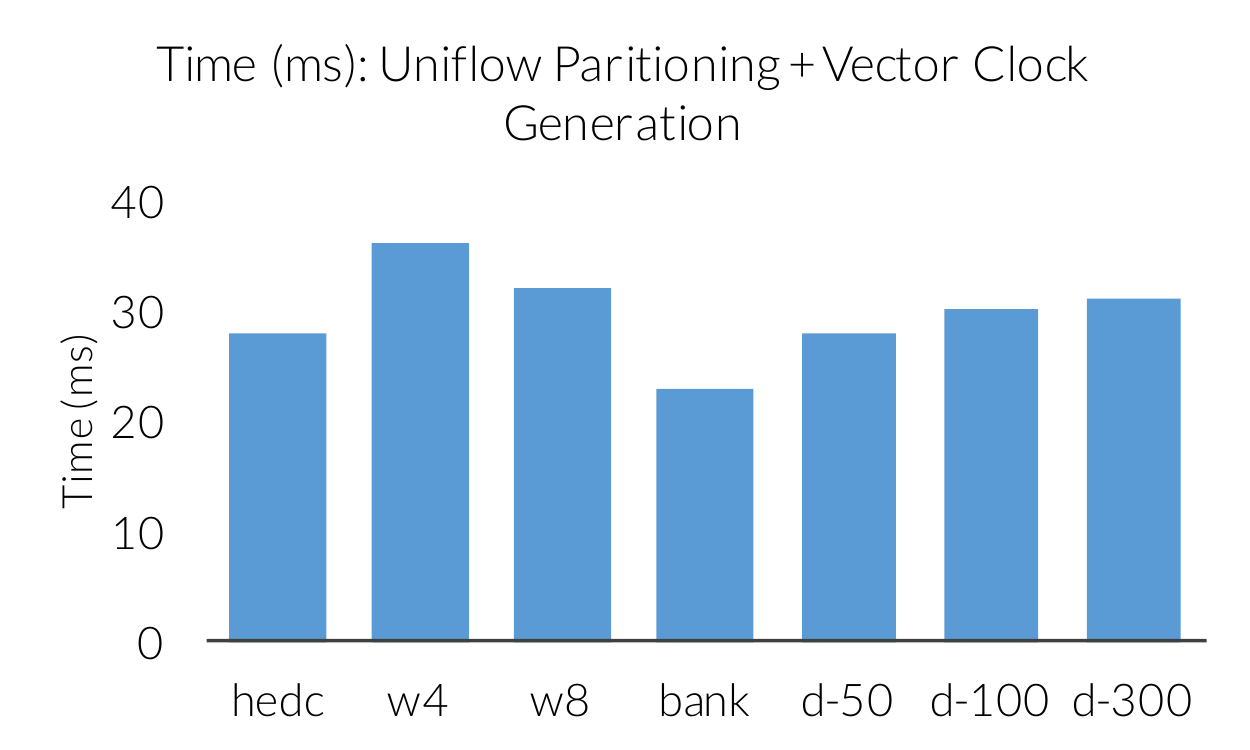}
\caption{Time taken to compute uniflow partition and corresponding vector clocks.} 
\label{fig:vcgen}
\end{figure}

We also report the memory consumed by the both of implementations during the traversal.  Note that 
the memory consumed for \our involves the memory taken by vector clocks created for uniflow chain 
partition, as well as vectors created during the traversal. For both the implementations, the memory consumption reported is for the 
traversal of the lattice,  and does not include the 
memory required to store the original poset.  
For the three benchmarks for which \bfs does not run out of memory, the comparison of overall memory consumed 
during the traversal is shown in Figure~\ref{fig:mem}. Figure~\ref{fig:ourmem} shows the memory 
consumption for \our on the remaining benchmarks. 

\begin{figure}
\centering
\subfloat[Comparison: memory consumed by \bfs and \our]{
\includegraphics[scale=0.5]{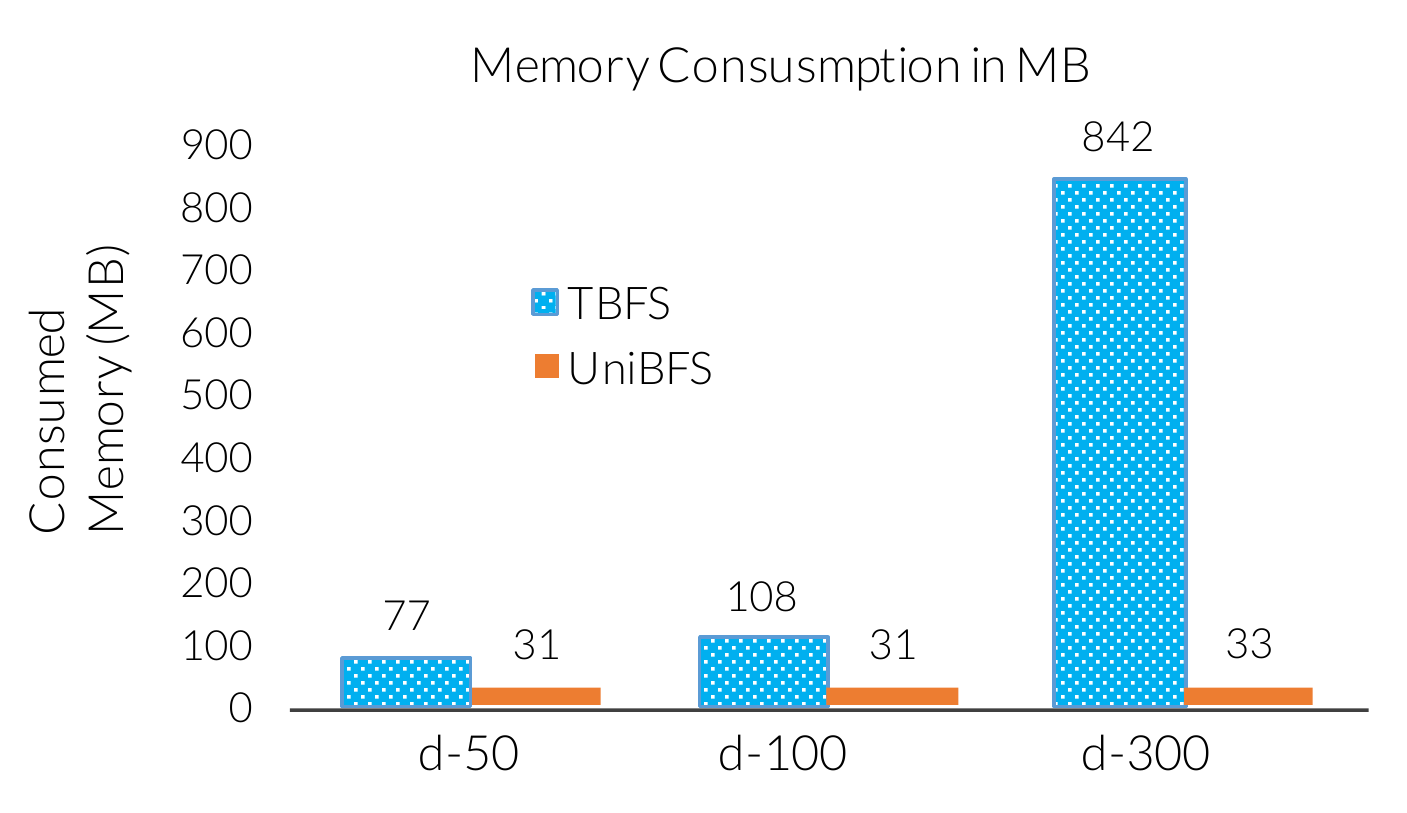}
\label{fig:mem}
}
\qquad
\subfloat[Memory consumption for \our for remaining benchmarks 
on which \bfs fails]{
\includegraphics[scale=0.5]{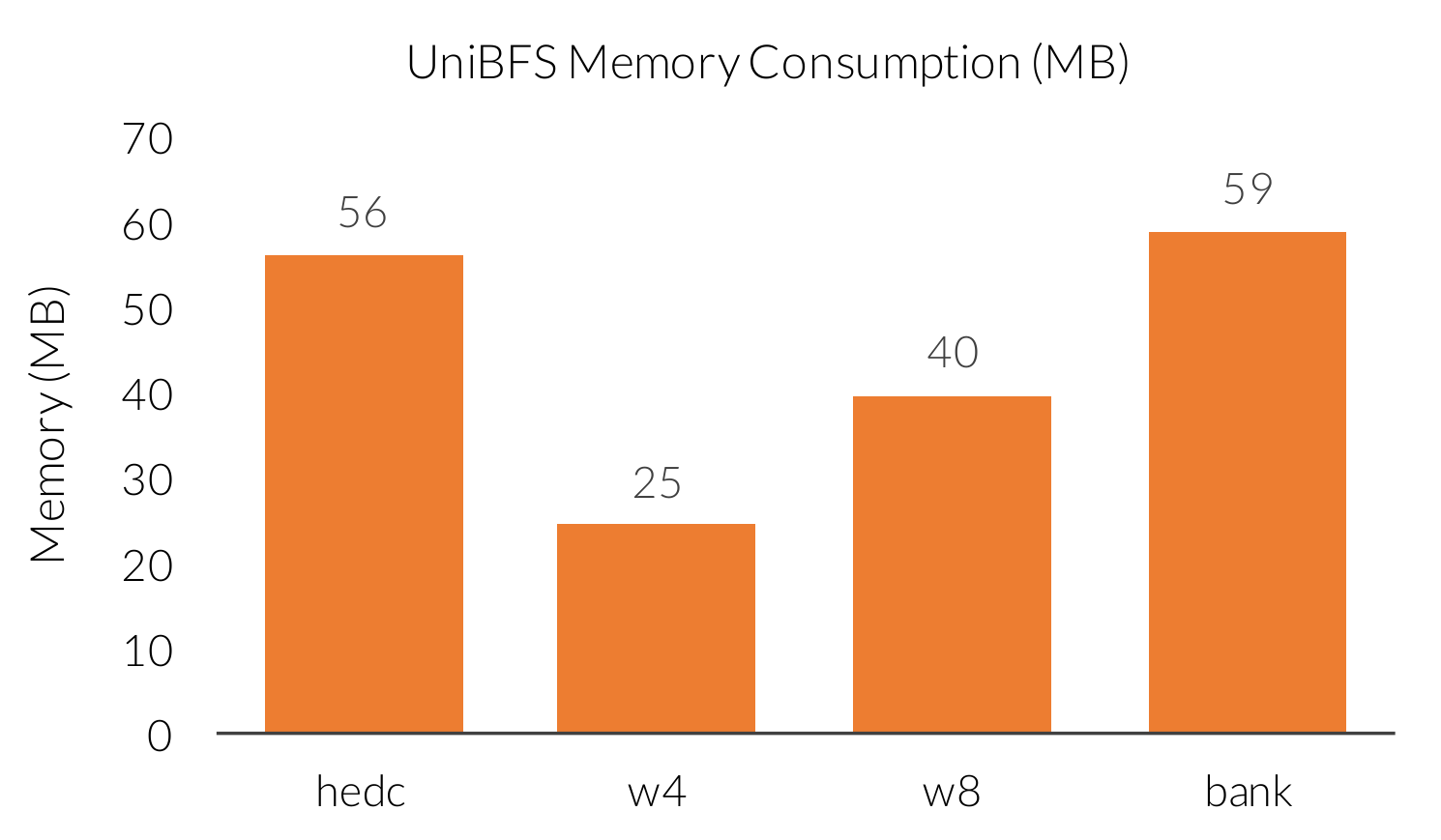}
\label{fig:ourmem}
}
\caption{Memory consumption for traversing all the consistent cuts}
\end{figure}

\begin{table}[]
\centering
\begin{tabular}{lrr}
\hline
Benchmark & \bfs time & \our time\\
\hline
d-50 & 200 ms & 44 ms\\
d-100 &  183 ms & 46 ms\\
w-4 & 4 ms & 35 ms\\
d-300 & 181 ms & 60 ms\\
bank & 24682 ms & 750 ms \\
hedc & 172 ms & 50 ms\\
w-8 & 15 ms  & 33 ms 
\end{tabular}
\caption{Runtimes for traversing consistent cuts of $rank = 30$.}
\label{tab:r25}
\end{table}

\begin{table}[]
\centering
\begin{tabular}{lrr}
\hline
Benchmark & \bfs time & \our time\\
\hline
d-50 & 242 ms & 25 ms\\
d-100 &  261 ms & 30 ms\\
w-4 & 14 ms & 36 ms\\
d-300 & 266 ms & 69 ms\\
bank & OOM* & 2828 ms \\
hedc & 2570 ms & 204 ms\\
w-8 & 101 ms  & 61 ms 
\end{tabular}
\caption{Runtimes for traversing consistent cuts of $rank = 50$. OOM* = out-of-memory error}
\label{tab:r50}
\end{table}

We now show that our algorithm can be particularly effective in performing 
traversals of a given rank.   In Table~\ref{tab:r25}, and Table~\ref{tab:r50} we report the runtimes of the 
implementations when we only are interested in traversing the consistent 
cuts of rank 30 (30 events have been executed in the system), 
and rank 50. 
for the seven benchmarks. When traversing the lattice for consistent cuts of rank 30 result, on the five out of the seven benchmarks --- except w-4 and w-8 --- our 
\our significantly outperforms \bfs  for traversing the consistent cuts of this given rank. 
For rank 50, we outperform \bfs on all but one, except w-4, of the benchmarks.  
}

\section{Future Work \& Conclusion}

Algorithm~\ref{alg:outline} can  perform the BFS traversal without regenerating 
the vector clocks for uniflow chain partitions. This is particularly 
beneficial for the 
 computations in which  
$|E| >> n$, and hence the $\bigo{|E|^2}$ space needed to regenerate 
the vector clocks is expensive. Observe that any chain partition, including a uniflow chain partition, 
of a computation is only an arrangement of its graph. Hence, we can implement Algorithm~\ref{alg:outline}
without regenerating new 
vector clocks, and by only finding the positions of the events in the 
uniflow chain partition. To do so, we assign a unique id to each event, 
and then place this event id on its corresponding uniflow chain. 
We also store a mapping of original vector clocks against the event ids. 
\remove{The steps of the traversal algorithm remain the same, only their interpretation 
is different now. Incrementing a cut by $K[i] = K[i] + 1$  now translates 
fetching the next event id, i.e. the id of $(i+1)$th event on chain $i$,  and
then merging its vector clock value to the current vector clock representation of $K$. 
 }
The space requirement for our algorithm will reduce to $\bigo{n_u \cdot n}$
as we do not regenerate vector clock, and computation of projections can be performed using $n_u \times n$ space 
instead of $n_u \times n_u$ space.
As a future work, we plan to implement and evaluate this strategy.   

It is easy to parallelize Algorithm~\ref{alg:outline}, as it 
traverses cuts of rank $r+1$ 
independently of those of rank $r$. We can perform a parallel traversal easily 
using a {\sf parallel-for} loop at line 3 of Algorithm~\ref{alg:outline}. We intend to 
implement this parallel approach and compare its performance against 
parallel traversal algorithms such as Paramount \cite{chang2015parallel}. 
\remove{
As of now, 
we do not know any polynomial time algorithm to find an optimal (smallest number of chains) uniflow chain 
partition of a poset. This remains an interesting and challenging open problem as the number 
of chains of the uniflow partition has direct impact on the runtime performance of our algorithm. 
}

 
The ubiquity of multicore and cloud computing has significantly
 increased the degree of parallelism in programs. 
 This change has in turn made 
verification and analysis of large parallel programs 
even more challenging.  
For such verification and analysis tasks, 
breadth-first-search based traversal of global states  
of parallel programs is a crucial routine.
We have reduced the space complexity of this routine from 
exponential to quadratic in the size of input computation.  
This reduction in space complexity allows us to analyze computation with high 
degree of parallelism with relatively small memory footprint ---  a task 
that is practically impossible with traditional BFS implementations.

\bibliographystyle{abbrv}
\bibliography{refs,citations,ppopp}  
%
%
\appendix
\section{Proof of Correctness}
\label{app:proofs}
\begin{l1}
\label{lem:min}
Let $G$ be any consistent cut of rank at most $r$. Then, 
$H =$~\getmin is the lexically 
smallest consistent cut of rank $r$ greater than or equal to $G$.
\end{l1}
\begin{proof}

We first show that $H$ is a consistent cut. 
Initially, 
$H$ is equal to $G$ which is a consistent cut.
We show that $H$ continues to be a consistent cut after every iteration of the
{\em for} loop. 
At  iteration $j$, we add elements from the $j^{th}$ chain from the bottom to $H$.
Since all elements from higher numbered chains are already part of $H$,
and all elements from lower numbered chains cannot be smaller than any of the
newly added element, we get that $H$ continues to be a
consistent cut.

By construction of our algorithm it is clear that rank of $H$ is exactly $r$.
We now show that $H$ is the lexically smallest consistent cut of rank $r$ greater than or equal to $G$.
Suppose
 not, and let $W <_l H$ be the lexically smallest consistent cut of rank $r$ greater than or equal to $G$.
Since $W <_l H$, let $k$ be the smallest index such that $W[k] < H[k]$.
Since $G \leq_l W$, $k$ is one of the indices for which we have added at least one event to $G$.
Because rank of $W$ equals rank of $H$, there must be an index $k'$ lower than $k$
such that $W[k'] > H[k']$. However, our algorithm 
forces that for $H$ for any index $k'$
lower than $k$, $H[k']$ equals $|P_{k'}|$. Hence, $W[k']$ cannot be greater than $H[k']$.
\end{proof}

\begin{l1}
Let $G$ be any consistent cut of rank at most $r$, 
 Then \getsucc
returns the least consistent cut of rank $r$ that is lexically greater than $G$.
\end{l1}
\begin{proof}
Let $W$ be the cut returned by \getsucc. We consider two cases.
Suppose that $W$ is null. This means that for all values of $i$, either
all elements in chain $P_i$ are already included in $G$, or on inclusion of the
next element in $P_i$, $z$, the smallest consistent cut that includes $z$ 
has rank greater than $r$. Hence, $G$ is lexically biggest consistent cut of rank $r$.

Now consider the case when $W$ is the consistent cut returned at line 16
 by {\sc GetMinCut}$(K ,r)$. We first observe that after executing line 11, $K$ is the next lexical 
consistent cut (of {\em any} rank) after $G$. If $rank(K)$ is at most $r$, then by Lemma~\ref{lem:min} we know that {\sc GetMinCut}$(K,r)$
 returns the smallest lexical consistent cut greater than or equal to $G$ of rank $r$.
 If $rank(K)$ is greater than $r$, then there is no consistent cut of rank $r$ such that
 $\forall k: i+1 \leq k \leq n_u: K[k] = G[k]$ and $K[i] > G[i]$ and $rank(K) \leq r$.
 Thus, at line 16 we use the largest possible value of $i$ for which
 there exists a lexically bigger consistent cut  than $G$ of rank $r$.
 \end{proof}

\section{Uniflow Partitioning Algorithm}
\label{sec:partition}

Lemma~\ref{lem:topo} establishes that every 
poset has a uniflow chain partition. The trivial 
 uniflow partition, however, will 
result in number of chains $n_u = |P|$. 
 Given that our traversal technique takes 
$\bigo{n_u^2}$ time per consistent cut in the worst case, it is beneficial 
to find a uniflow chain partition that has fewer number of chains. 
We now discuss an online chain partitioning  algorithm to find such non-trivial uniflow chain partitions 
of posets. 
\remove{
In the interest of space, we present an offline algorithm that processes 
the full poset of a computation in Appendix~\ref{app:offlinepart}. }

\subsection{Online Algorithm} 

This algorithm processes events of the 
computation $P = (E,\ra)$ in an online manner: when a process 
$P_i$ executes event $e$ it sends the event information to 
this partitioning algorithm. 
Algorithm
~\ref{alg:onlinepart} shows the steps of finding an appropriate 
chain for $e$ in the uniflow partition.  
\begin{algorithm}[tbh]
\caption{{\sc FindUniflowChain}($e$)}
\label{alg:onlinepart}
\begin{algorithmic}[1]
\Require An event $e$ of the computation $P = (E, \ra)$
on $n$ processes 
\Ensure $e$ is placed on a chain in the uniflow chain 
partition of $P$ 

\State {$maxid$: id of highest uniflow chain till now}
\State {$uid = e.procid$} \Comment{\textcolor{blue}{start with chain that executed $e$}}

\For{each causal dependency $dep$ of $e$}
   \State {$uid = $ MAX$(dep.procid,uid)$}
\EndFor
\State {\textcolor{blue}{//now check if there exists any chain with the same id}}
\If {$\exists$ a chain with $id = uid$}
   \State {$f = $~last event on this chain}
   \If {$e || f$} \Comment {\textcolor{blue}{$e$ is concurrent with $f$}  }
      \State {$maxid++$} \Comment{\textcolor{blue}{increment max used chain id}}
      \State {create new chain with id $= maxid$}
      \State {add $e$ at the end of this chain}
   \Else{} \Comment {\textcolor{blue}{$e$ not concurrent with $f$}}
   \State{add $e$ at the end of chain (with $id = uid$)}
   \EndIf
\Else
      \State {create new chain with id $uid$}
      \State {add $e$ at the end of this chain}
      \State {$maxid = uid$} \Comment {\textcolor{blue}{increment max assigned chain id}}
\EndIf
\end{algorithmic}
\end{algorithm}

 Given an event $e$, we set 
its uniflow chain, $uid$, to the id of the chain (process) 
on which it was executed. Then, we go over 
all its causal dependencies, and in case any of 
the dependencies were placed on higher numbered 
chains, we update the $uid$ (lines 3--4). 
We know that to maintain the uniflow chain partitioning,
$e$ must be placed either on a chain 
with id $uid$, or above it. Lines 6--8 check if 
there already exists a chain with that id, and 
if the last event on this chain is concurrent with $e$. 
If so, we cannot place $e$ on this chain and must 
put it on a chain above --- possibly by creating a 
new chain (lines 9--11). Otherwise, we can place 
$e$ at the end of this chain, and do so at 
line 12. If no chain has been created with $uid$ as 
its number, that means $e$ is the first event 
on some chain (process) in $P$ and we must 
create a new chain in our uniflow  partition for it. 
This is done in lines 15--17.

Let us illustrate the execution of the algorithm on the 
poset of \figref{fig:transformvc}(a).  Let 
$\mu$ denote the uniflow chain partition, which is initially empty. 
Suppose the first event on process 
$P_1$ is the very first event sent to the this algorithm.
As there 
is no event in $\mu$, this event will be placed 
on chain id $1$. As we assume an online setting, the first event on $P_2$ must 
going to be presented next.
As this event also has no causal dependencies, the $uid$ value for it at line 6 will be $2$ --- the 
id of the process that executed it.  
As there is no chain with id $2$, we execute lines 14--17 to create a new chain
and place this event on chain $2$ in $\mu$. Suppose the next event 
to arrive is the second event on $P_2$. 
Even though is causally dependent on the first event on $P_1$, 
its $uid$ value after the loop of line 3--4 is still 2 as we take the 
maximum of the ids. 
 There 
is a chain with id $2$ in $\mu$, and this 
event is not concurrent with the last event on this chain. Hence, 
we execute lines 12--13, and place this event as the last event on chain $2$.
The last event to arrive will be the second event on $P_1$. After executing 
lines 3--4, the $uid$ value for this event will be 2. As there is 
a chain with id 2 in $\mu$, we will compare this event with the last event 
on chain 2. However, the last event on chain 2 and this event are concurrent (line 8). 
Hence, we will be forced to create a new chain, with id 3, and place the 
event on this chain as
per lines 9--11.  

Let $\Delta$ be the maximum in-degree of any event in the computation. 
Then lines 3--4 of 
 Algorithm
~\ref{alg:onlinepart} perform $\bigo{\Delta}$ work. Checking for existence of a 
chain id at line 6 is a constant time operation as we use a hash-table for storing 
the chains against their ids. The check for concurrency of two events 
is $\bigo{n}$ as we can use the original  vector clocks of the two events. 
Lines 9--11 then perform constant work. If the events are not concurrent, 
and we execute line 13, we still perform constant work in appending the event 
at the end of a chain. Lines 15--17 also perform constant work.  
Hence, the total work performed by the algorithm is 
$\bigo{n + \Delta}$ per event.  

\section{Optimized Implementation of \getsucc}
\label{app:getsuccopt}
\begin{algorithm}[tbh]
\caption{{\sc GetSuccessorOptimized}($G,r$)}
\label{alg:succopt}
\begin{algorithmic}[1]
\Require $G$: a consistent cut of rank $r$ 
\Ensure $K$: lexical successor of $G$ with rank $r$ 

\State {\Call{ComputeProjections}{$G$}} \Comment{\textcolor{blue}{$G$'s projections}}
\State {$K = G$} \Comment{\textcolor{blue}{Create a copy of $G$ in $K$}}
\For{($i = 2; i \leq n_u; i{+}{+}$)} 
\If{next element on $P_i$ exists}
   \State $K[i] = K[i] + 1$\Comment{\textcolor{blue}{increment cut in $P_i$}}

\State {\textcolor{blue}{//fix dependencies using projections}}
\State {$vc = $ vector clock of event number $K[i]$ on $P_i$}
\State {\textcolor{blue}{//take component-wise max}}
\For{$(k = i-1; k > 0; k--)$}
  \State {$K[k] = $ MAX($vc[k]$, $proj[i][k]$)}
\EndFor

\If{ $rank(K) \leq r$} \State {\Return \Call{GetMinCut}{$K,r$}} \Comment{\textcolor{blue}{make  
$K$'s rank equal to $r$}}
\EndIf
\EndIf
\EndFor
\State {\Return null} \Comment{\textcolor{blue}{could not find a candidate cut}}
\end{algorithmic}
\end{algorithm}

\section{Memory Consumption for Traversing Specific Ranks}
\label{app:memory32}
Table~\ref{tab:memory32} shows the memory consumed by the 
three algorithms (traditional BFS, Lex, and Uniflow based BFS)
in traversing consistent cuts of specific ranks, as well as 
all cuts of rank less than or equal to 32. 
\begin{table}[tbh]
\centering
\caption{Heap Memory Consumed (in MB) for {\sf tbfs}: Traditional BFS, {\sf lex}: Lexical, and {\sf uni}: Uniflow BFS implementations to traverse cuts of given ranks. \textcolor{red}{$\times$}= out-of-memory error}
\label{tab:memory32}
\begin{tabular*}{\textwidth}{@{\extracolsep{\fill}}l|rrr|rrr|rrr|rrr@{}}
\toprule
\multicolumn{1}{c}{Name} & \multicolumn{3}{c}{$r =\frac{|E|}{4}$}                                          & \multicolumn{3}{c}{$r =\frac{|E|}{2}$} & \multicolumn{3}{c}{$r =\frac{3|E|}{4}$} & \multicolumn{3}{c}{$r \leq 32$} \\ \midrule
                         & {\sf tbfs}                    & {\sf lex}                    & {\sf uni}                    & {\sf tbfs}       & {\sf lex}      & {\sf uni}      & {\sf tbfs}       & {\sf lex}       & {\sf uni} & {\sf tbfs}       & {\sf lex}       & {\sf uni}      \\ 
d-100    & 95   & 32   & 41  & 121 & 29   & 41    & 134 & 32   & 42  & 112   & 32   & 42  \\
d-300    & 107   & 33   & 53  & 342 & 32   & 54    & 583  & 32    & 54  & 113   & 31   & 42  \\
d-500    & 299  & 33   & 56 & 695 & 32  & 55  & 1604 & 34  & 55  & 112   & 32   & 41  \\
bank     & 1014  & 21  & 52 & \textcolor{red}{{$\times$}}  & 22  & 54  &\textcolor{red}{{$\times$}}   & 21  & 54  & 1312 & 22  & 54\\
hedc     & 934  & 33 & 61 & \textcolor{red}{{$\times$}} & 34 & 62 & \textcolor{red}{{$\times$}}   & 34 & 62  & 602  & 31 &    60\\
w-4      & 83    & 21  & 49  & 313  & 22    & 49   & 301  & 21    & 51   & 36     & 20   & 49 \\
w-8      & 1786 &  27 & 44 & \textcolor{red}{{$\times$}} & 28 & 43 & \textcolor{red}{{$\times$}}   & 28 & 45 & 1240    & 28 & 43 \\
\bottomrule
\end{tabular*}
\end{table}

\end{document}